\documentclass[fleqn,usenatbib]{mnras}

\usepackage[T1]{fontenc}

\DeclareRobustCommand{\VAN}[3]{#2}
\let\VANthebibliography\thebibliography
\def\thebibliography{\DeclareRobustCommand{\VAN}[3]{##3}\VANthebibliography}

\usepackage{graphicx}	
\usepackage{amsmath}	
\usepackage{amssymb}	

\newcommand{\msun}{{\,\rm M_\odot}}
\newcommand{\kms}{\,{\rm km}\,{\rm s}^{-1}}
\def\gsim{ \lower .75ex \hbox{$\sim$} \llap{\raise .27ex \hbox{$>$}} }
\def\lsim{ \lower .75ex \hbox{$\sim$} \llap{\raise .27ex \hbox{$<$}} }

\usepackage{newtxtext,newtxmath}

\title[DM and satellite spatial distributions]{The spatial distribution of Milky Way satellites, gaps in streams  and the nature of dark matter}

\author[M. R. Lovell et al.]{
Mark R. Lovell$^{1}$\thanks{E-mail: lovell@hi.is},
Marius Cautun$^{2}$, Carlos S. Frenk$^{3}$, Wojciech A. Hellwing$^{4}$ and Oliver Newton$^{5}$
\\
$^{1}$Center for Astrophysics and Cosmology, Science Institute,  University of Iceland, Dunhaga 5, 107 Reykjav\'ik, Iceland \\
$^{2}$Leiden Observatory, Leiden University, PO Box 9513, NL-2300 RA Leiden, the Netherlands \\
$^{3}$Department of Physics, Institute for Computational Cosmology, University of Durham, South Road, Durham, DH1 3LE, UK\\
$^{4}$Center for Theoretical Physics, Polish Academy of Sciences, Al. Lotnik\'ow 32/46, 02-668 Warsaw, Poland \\
$^{5}$Univ. Lyon, Univ. Claude Bernard Lyon 1, CNRS, IP2I Lyon/IN2P3, IMR 5822, F-69622, Villeurbanne, France}

\date{Accepted 2021 August 20. Received 2021 July 21; in original form 2021 April 14.}

\pubyear{2021}

\begin{document}
\label{firstpage}
\pagerange{\pageref{firstpage}--\pageref{lastpage}}
\maketitle

\begin{abstract}
The spatial distribution of Milky Way (MW) subhaloes provides an important set of observables for testing cosmological models. These include the radial distribution of luminous satellites, planar configurations, and the abundance of dark subhaloes whose existence or absence is key to distinguishing  amongst dark matter models. We use the {\sc COCO} $N$-body simulations of cold dark matter (CDM) and 3.3~keV thermal relic warm dark matter (WDM) to predict the satellite spatial distribution in the limit that the impact of baryonic physics is minimal. We demonstrate that the radial distributions of CDM and 3.3~keV-WDM luminous satellites are identical if the minimum pre-infall halo mass to form a galaxy is $>10^{8.5}$~$\msun$. The distribution of dark subhaloes is significantly more concentrated in WDM due to the absence of low mass, recently accreted substructures that typically inhabit the outer parts of a MW halo in CDM. We show that subhaloes of mass $[10^{7},10^{8}]$~$\msun$ and within 30~kpc of the centre are the stripped remnants of larger haloes in both models. Therefore their abundance in WDM is $3\times$ higher than one would anticipate from the overall WDM subhalo population. We estimate that differences between CDM and WDM concentration--mass relations can be probed for subhalo--stream impact parameters $<2$~kpc. Finally, we find that the impact of WDM on planes of satellites is likely negligible. Comprehensive comparisons with observations will require further work with high resolution, self-consistent hydrodynamical simulations.         
\end{abstract}




\section{Introduction}

The distribution of galaxies is a key observable that a viable cosmological model must reproduce in order to be an accurate description of our Universe. The success of the $\Lambda$ cold dark matter ($\Lambda$CDM) model in explaining the properties of the large-scale Universe is remarkable, including the prediction of the existence of baryonic acoustic oscillations (BAOs) \citep{cole05,Eisenstein05}. The predictions for galaxy distributions on small scales are controversial, with numerous  claimed discrepancies between observations and theoretical predictions \citep[see ][ for a review;]{bullock2017}, these discrepancies could be removed by baryon physics, dark matter physics, or a combination of the two. 

In this paper we will consider three key observables related to the spatial distribution of satellite galaxies in the context of dark matter studies: the radial distribution of the Milky Way (MW) luminous satellite galaxies, the radial distribution of dark satellites, and the apparent distribution of some galaxies in planar structures.

First, we discuss the radial distribution of MW satellites. This distribution is known to be more concentrated than the distribution of massive subhaloes in many MW-analogue CDM simulations. The reasons for this discrepancy could include that the timescale to sink to the centre of the host is anticorrelated with the mass of the galaxy at infall, with more massive subhaloes more likely to host satellite galaxies ; thus much of this discrepancy disappears when predictions are made for luminous satellites as opposed to simply massive subhaloes \citep{Font11,Lovell17b,Newton18,Bose20,Samuel20}. It is likely that our census of massive satellites at distances $>100$~kpc from the MW is incomplete, in which case the satellite distribution would be less concentrated than currently thought; that the distribution of M31 satellites is consistent with both $N$-body and hydrodynamical simulation predictions supports this hypothesis \citep{Yniguez14,Font21}. There is also the possibility that the accretion of the Magellanic Clouds has introduced an exceptionally large number of dwarf galaxies on orbits that run atypically close to the MW centre \citep{SantosSantos20b}.

A more speculative, and certainly interesting problem for the purposes of understanding the particle nature of dark matter, is the distribution of dark satellites: subhaloes that have had all of their baryons evaporated by reionisation radiation before they could form any stars \citep{Bullock_00,Benson_02}. The CDM model predicts the existence of such subhaloes down to scales of $1$~$\msun$ or even smaller \citep{Schneider13}. The presence of such subhaloes has been the subject of study in two contexts: the lensing of background galaxies and quasars by elliptical galaxies hosted in $\sim10^{13}$~$\msun$ haloes, where some detections have been reported \citep{Vegetti10,Hezaveh16} and, more pertinently for this paper, stellar streams around the MW where interactions of dark subhaloes may have ripped gaps in stellar streams \citep{Yoon11,Carlberg13,Erkal15,Bonaca19}. Detections of individual gaps have been reported, although it is not clear whether such gaps are indeed due to dark matter subhaloes or instead to some denser object \citep{Amorisco16,Bonaca19}.  

Finally, the apparent location of some MW satellites in a planar structure is a particularly puzzling discrepancy. There is considerable evidence that some luminous satellites are located in a flat plane \citep{LyndenBell76,Libeskind05,Metz08,Metz09,Lux10,Pawlowski20}. Whether or not such planes are rotationally coherent is much more difficult to either confirm or refute given the challenges of obtaining proper motions for satellites with sufficient precision \citep{SantosSantos20} and due to the satellites following complex orbits that are affected by the non-spherical shape of the host halo and other massive satellites \citep{Shao19}. The likelihood of having a MW-like plane of satellites is still an open debate, ranging from ~0.1~per~cent \citep{Pawlowski14,Shao19} to as high as 10~per~cent when accounting for the diversity of potential $\Lambda$CDM planes of satellites and the look-elsewhere effect \citep{Cautun15}. In CDM, planes of satellites arise from the anisotropic accretion of satellites, including group infall and filamentary accretion \citep{Kang05,Zentner05,Li08,Libeskind05,Libeksind07,Lovell2011,Shao18} and chance alignments \citep{Cautun15}; the destruction of satellites on radial orbits by the disc may also play a role, although not to a sufficient degree to alleviate the tension with observations \citep{Ahmed17}.

To a first approximation, the radial distribution of satellites is determined by fluctuations on a $\sim1$~Mpc scale, which are the same in both CDM and WDM with thermal relic masses $>2$~keV. However, there are important secondary effects that can lead to differences, especially for subhaloes whose mass is close to the WDM cutoff in the power spectrum. The primary difference in halo properties between CDM and WDM models is that in the latter, the overall abundance of structure and therefore the number of dwarf haloes is suppressed.  A second effect is the change in the internal properties of WDM haloes close to the mass cutoff scale, which are characterized by lower densities \citep{Colin00,Lovell12} and thus can potentially experience faster mass loss and disruption when orbiting around a more massive host such as the MW. However, it has been shown that the switch from one model to the other also affects the radial distribution of low mass subhaloes, both in MW-analogue haloes \citep{Lovell14,Bose17a} and in $\sim10^{13}$~$\msun$ haloes that are used for lensing constraints \citep{Despali19b}. In both cases, the radial distribution of subhaloes with masses below the WDM characteristic half-mode mass, $M_\rmn{hm}$, is steeper in WDM than in CDM, which can be seen as a change in the radial distribution of satellite galaxies if the half-mode mass is high enough. As we discussed above, the analysis of gaps in streams is a popular test of WDM \citep[e.g.][]{Banik19b,Benito20,Banik21}, and it is important to test whether any differences in radial distribution will either compound or compensate for the impact of the pure suppression of subhalo abundance in WDM. 

Finally, one can envisage a situation in which the nature of the dark matter can affect the presence of satellite planes. The phase space distribution of satellites is influenced by group infall \citep{Li08,Shao18} and filamentary accretion \citep{Libeskind05,Lovell2011}. Haloes that form in low density regions, such as lower mass filaments, are delayed in their collapse times, such that their gas is photo-evaporated by reionisation before collapse and they can no longer form stars; they will also be less massive. \citet{Lovell19a} showed that the introduction of a power spectrum cutoff makes this delay longer than is the case for CDM, in a manner that is anti-correlated with the local density. If the WDM model is sufficiently extreme, the formation of galaxy-mass haloes in low mass filaments may be inhibited, restricting infall of dwarf galaxies to a small number of massive filaments and thus a greater proportion of dwarf galaxies will be members of planes. The crucial question is whether the cutoff scale at which this effect happens is above or below the cutoff scale that is consistent with current bounds on WDM \citep{Enzi21,Nadler21,Newton21}; this problem has not previously been addressed.

In this study we use the COCO $N$-body simulations \citep{Hellwing16,Bose16a} to examine the spatial configurations of satellites in the WDM and CDM cosmologies. We will extract MW-analogue haloes that are produced in both simulations, and analyse the infall history of their subhaloes. We will briefly discuss the radial distribution of subhaloes as pertains to dwarf galaxies, devote the bulk of our analysis to the distribution and concentration of dark subhaloes, and present a brief coda on the question of planes of satellites. 

A definitive study would also account for the contraction of the host halo and the presence of the stellar disc in destroying subhaloes on radial orbits  \citep{Blumenthal86,Gnedin04,Ahmed17,GarrisonKimmel17,Sawala17,Richings20}. This would require numerous hydrodynamical simulations and is beyond the scope of this paper: we will therefore highlight the limitations of our $N$-body approach throughout the text as necessary, including in a dedicated subsection (Section~\ref{subsec:barimpact}).

This paper is organised as follows. We describe our simulations in Section~\ref{sec:sims}, present our results in Section~\ref{sec:res}, and draw conclusions in Section~\ref{sec:conc}. 

\section{Simulations}
\label{sec:sims}

The simulations used in this paper are the COCO $N$-body cosmological simulations. These are cosmological zoomed simulations of a high resolution sphere, $\sim24.7$~Mpc in radius, embedded within a periodic cube of side-length 100~Mpc. The high resolution simulation particle mass is $1.6\times10^{5}$~$\msun$, the gravitational softening length is $\epsilon=0.33$~kpc, and the cosmological parameters are consistent with the WMAP7 cosmology \citep{wmap11}: Hubble parameter $h=0.704$, matter density $\Omega_{0}=0.272$, dark energy density, $\Omega_{\Lambda}= 0.728$; spectral index, $n_{s}=0.967$; and power spectrum normalization $\sigma_{8}= 0.81$. The simulations were performed with the {\sc p-gadget3} galaxy formation code, which is an updated version of the publicly available {\sc gadget2} code \citep{Springel_05}. Haloes are identified with the friends-of-friends algorithm, and are deconstructed into subhaloes with the {\sc subfind} halo finder \citep{Springel01}. The minimum number of particles required to identify a subhalo is 20, which is a mass of $3.2\times10^{6}$~$\msun$.

Two copies of the COCO volume were simulated: a CDM version \citep{Hellwing16} followed by a WDM counterpart \citep{Bose16a}. The WDM model is for a thermal relic with particle mass 3.3~keV, and has been implemented using the transfer function of \citet{Viel05}. The model can also be described by the half-mode mass, $M_\rmn{hm}$, which is defined as the mass corresponding to the sharp $k$-space filter where the amplitude of the WDM transfer function falls to half of the CDM transfer function. For this 3.3~keV thermal relic, $M_\rmn{hm}=2.8\times10^{8}$~$\msun$, which is approximately the same $M_\rmn{hm}$ as for a resonantly-produced sterile neutrino \citep{Asaka:06c,Laine08,Lovell16}, of mass of 7~keV and lepton asymmetry, $L_{6}=10$ \citep{Lovell20}. WDM subhaloes that originate from spurious fragmentation of filaments are removed according to the method presented in \citet{Lovell14}. Briefly, we identify the shapes of haloes in the initial conditions -- otherwise known as the shapes of the `protohaloes' -- and exclude those for which the protohalo sphericity is lower than 0.165. We also remove haloes for which the peak mass over the history of the simulation is lower than half the limiting mass, $M_\rmn{lim}$, which is a function of the WDM power spectrum and the simulation resolution: for the COCO volume this mass cutoff takes the value $\sim3.8\times10^{7}$~$\msun$.

We now discuss our strategy and procedure for selecting host haloes. Our goal is to select haloes that are of a similar mass to the Milky Way, that are relaxed systems, and that are isolated. The first criterion is based on virial mass, for which we use $M_{200}$. This is the mass enclosed inside a radius within which the density is 200 times the critical density required for collapse; this radius is labelled $r_{200}$. For our halo relaxation criterion, we use the ratio of the mass of the central smooth halo to the total mass gravitationally bound to the halo as measured by {\sc subfind}; see \citet{Neto07} for a discussion on halo relaxation criteria. Third, we introduce an isolation criterion by requiring that a candidate halo is a minimum distance from any other halo of $M_{200}>1\times10^{12}$~$\msun$.  We identify haloes in the CDM halo catalogue that have a virial mass, $M_{200}=[1.0,1.5]\times10^{12}\msun$, a smooth-to-total mass ratio $>0.8$, and a minimum separation from other $M_{200}>1\times10^{12}$~$\msun$ haloes of 2~Mpc. Note that this last criterion excludes Local Group-like systems given the MW-M31 distance is $\sim$0.75~Mpc. We then use the halo position and $M_{200}$ to find matches to these haloes in the WDM simulation; our final sample consists of 24 hosts. We have checked these matches by comparing the density profiles of WDM and CDM hosts, and find that they agree to better than 5~per~cent. We therefore highlight that massive haloes have the same profiles in WDM as in CDM. 

\section{Results}
\label{sec:res}

We present our results in five subsections. We first discuss the radial distribution of satellites in general (Section~\ref{subsec:rd}), and then in a simple application to gaps-in-streams physics (Section~\ref{subsec:isp}). We examine the mass-concentration relations of subhaloes in Section~\ref{subsec:sd}, and in Section~\ref{subsec:barimpact} discuss the potential impact of baryonic processes and the caveats these impose on our predictions for subhalo properties in the host inner regions. We end our results presentation with a discussion on planes of satellites in CDM and WDM in Section~\ref{subsec:pos}.

\subsection{Radial distributions}
\label{subsec:rd}

\begin{figure}
    \centering
    \includegraphics[scale=0.42]{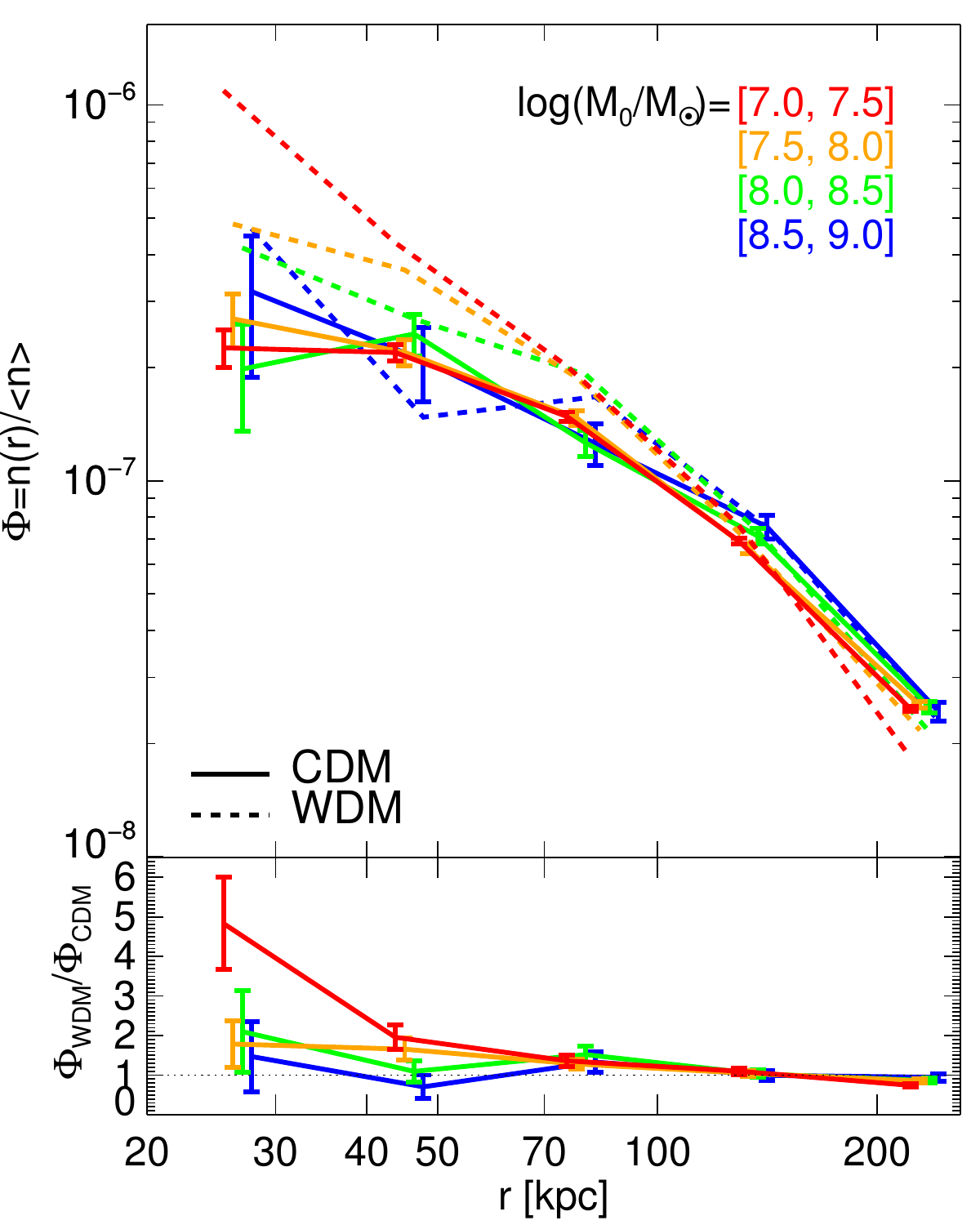}
    \caption{Top panel: the radial distribution of subhaloes in MW halo-analogues, normalised by the number of subhaloes in each mass bin. CDM data are represented with solid curves and WDM data with dashed curves. The curve colour associated with each mass bin is given in the figure legend. Error bars are shown for CDM only. Bottom panel, the ratio of the WDM and CDM radial distributions. The error bars are the propagation of Poisson errors on both data sets.}
    \label{fig:RD0}
\end{figure}

We begin by computing the radial distribution of our MW analogue satellites. We separate our satellites into four bins by $z=0$ subhalo mass, which we define as the mass gravitationally bound to the satellite as computed by {\sc subfind}; we denote this mass as $M_{0}$. We then stack the subhaloes from all hosts into a single set of radial distributions. We present our results for CDM and WDM in Fig.~\ref{fig:RD0}.

The radial distributions of satellites more massive than $10^{8.5}$~$\msun$ are largely the same in the two models. At lower masses -- which correspond to subhaloes less massive than $3\times10^{8}$~$\msun$, i.e. $M_\rmn{hm}$ -- a strong discrepancy appears. For example, the normalised radial density of $M_{0}=[10^{7.5},10^{8.0}]$~$\msun$ WDM subhaloes at 30~kpc is twice that in CDM at radii $<70$~kpc, and in the $[10^{7.0},10^{7.5}]$ bin the normalised abundance is a factor of five times the CDM equivalent.\footnote{Note that the {\it total} number of WDM subhaloes in this mass bracket and radius range is still smaller than the number of CDM subhaloes in these same brackets: it is only the density normalised to the number of all WDM subhaloes that is larger.} We thus recover the intriguing results found in \citet{Bose17a} and \citet{Despali19b}. 

Crucially, the difference between CDM and WDM correlates with halo mass, with bigger discrepancies between the two for mass bins $M_{0}<M_\rmn{hm}$. The population of subhaloes in a particular mass bin is composed of two broad classes of objects. First, there are low mass haloes that were accreted and have experienced very little, if any, stripping. The second class are haloes that were much more massive at accretion and have since lost mass to tidal stripping. The relative number of these haloes will change between CDM and WDM, but it is less clear whether the suppression in WDM subhalo densities influences the process of dynamical friction. We first check for differences in subhalo processing, and then consider the difference between the contributions of stripped high mass and unstripped low mass subhaloes to each $M_{0}$ bin.

\citet{Chandrasekhar43,Lacey93,Simha17} showed that the stripping time of the satellite becomes smaller as the mass ratio between host and satellite tends to 1. If the host halo density profiles were the same in the two models, we would expect that CDM and WDM subhaloes of the same infall mass should have the same dynamical friction time and therefore not contribute to the discrepancy in radial profiles. 

\begin{figure}
    \centering
    \includegraphics[scale=0.42]{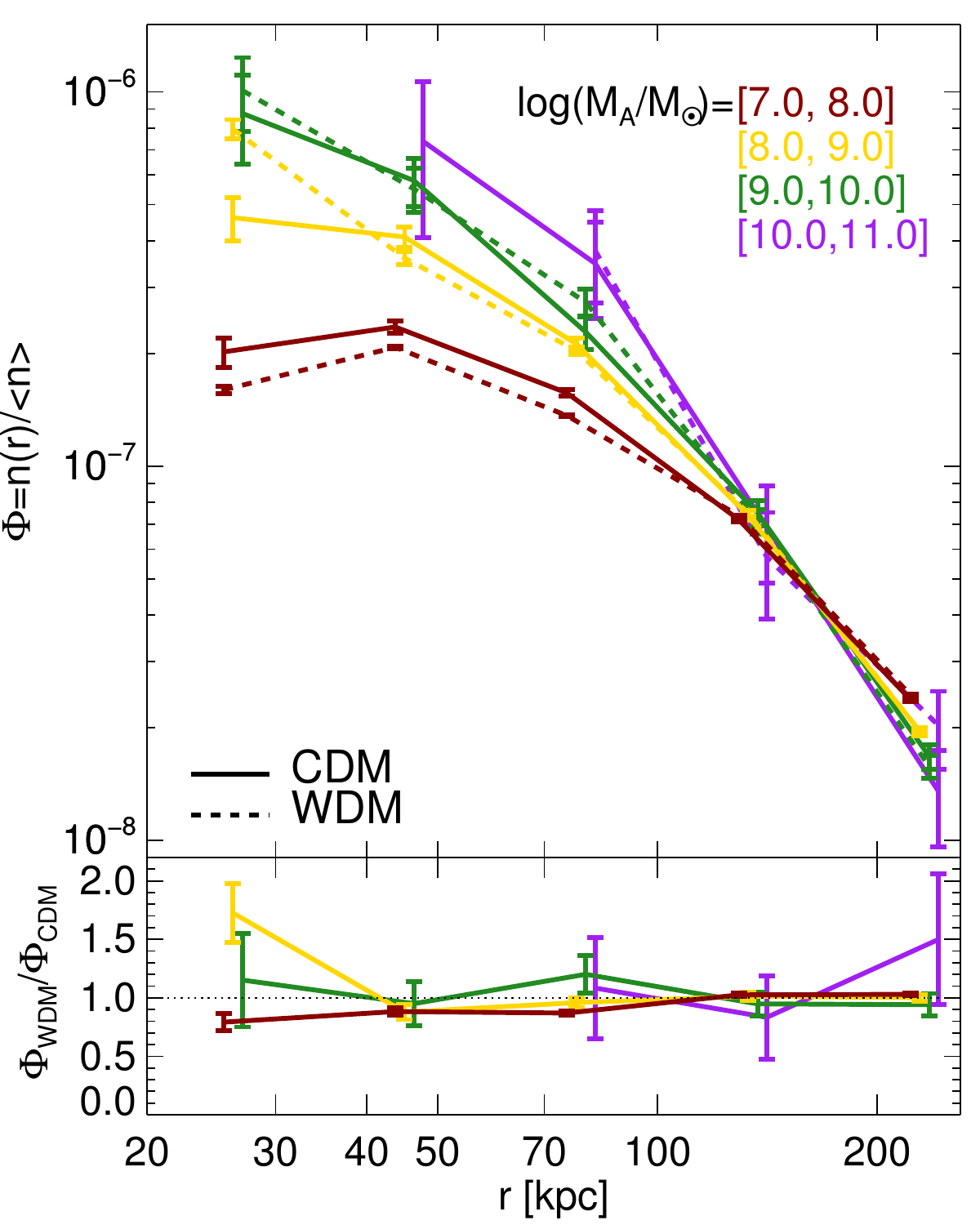}
    \caption{Top panel: the radial density profile of subhaloes binned by accretion mass, $M_\rmn{A}$. CDM curves are shown as solid lines and WDM curves as dashed lines. The relationship between curve colour and $M_\rmn{A}$ bin is given in the figure legend. Error bars are Poisson. Bottom panel: the ratio of the two data sets.}
    \label{fig:RDA}
\end{figure}

We first check the assertion that the model does not impact the host density profile. We computed density profiles for each of the 24 pairs of hosts, and calculated the ratio of WDM-to-CDM densities: we found that the median WDM-to-CDM profile enhancement at any radius is 3~per~cent, and 84~per~cent of hosts show a WDM-to-CDM enhancement of less than 10~per~cent at $r>20$~kpc. We then check the hypothesis that dynamical friction is similar for CDM and WDM subhaloes by computing the radial distributions of subhaloes binned by the gravitationally bound mass at accretion, $M_\rmn{A}$, instead of $M_{0}$. We define the time of accretion as the first time that the subhalo passes within $r_{200}$ of its $z=0$ host; for subhaloes that are within 300~kpc but never enter $r_{200}$, $M_\rmn{A}=M_{0}$. We plot the result in Fig.~\ref{fig:RDA}. Note that we use a different set of mass bins to Fig.~\ref{fig:RD0}.

There is a sharp distinction between the different $M_\rmn{A}$ bins. The $M_\rmn{A}=[10^{10},10^{11}]$~$\msun$ subhalo radial distribution is steeper than the $M_\rmn{0}=[10^{9.5},10^{10}]$~$\msun$ curve, and decreasing the bin mass results in a progressively shallower radial distribution \citep[see ][]{HanJ16}. By contrast, there is no clear evidence for a systematic deviation between the CDM and WDM cases, which implies that satellites of a given $M_\rmn{A}$ experience a similar degree of dynamical friction regardless of whether they are WDM or CDM. There is an apparent difference at radii $<40$~kpc in the $[10^{8},10^{9}]$~$\msun$ bin, with WDM subhaloes 60~per~cent more abundant than in CDM, but the effect shrinks to 20~per~cent at the lowest mass bin. We have repeated this process with the peak mass, $M_\rmn{P}$, which we define as the peak mass obtained by each subhalo across its merger tree main branch, and we find the results are practically identical as for $M_\rmn{A}$.  We note that subhaloes can survive more easily in the absence of a stellar disc: if a disc were to be included, the distributions would become less concentrated. Therefore our results can be considered to be an upper bound on the concentration of the satellite distribution.

\begin{figure*}
    \centering
        \setbox1=\hbox{\includegraphics[scale=0.23]{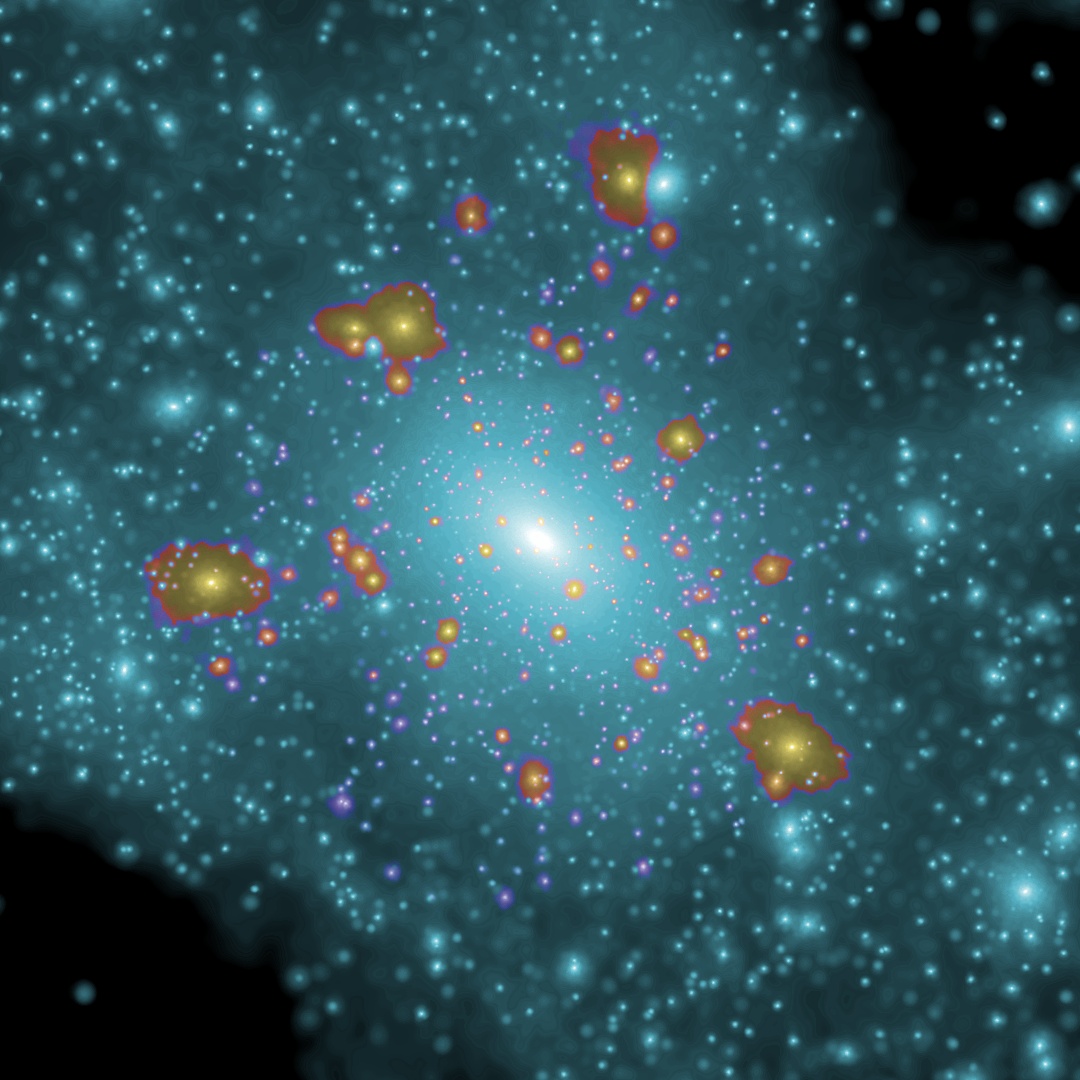}}
    
     \includegraphics[scale=0.23]{COCO_CDM_153_Bsize300kpc_Halo006.jpg}\llap{\makebox[\wd1][l]{\raisebox{0.7\wd1}{\includegraphics[width=0.36\textwidth]{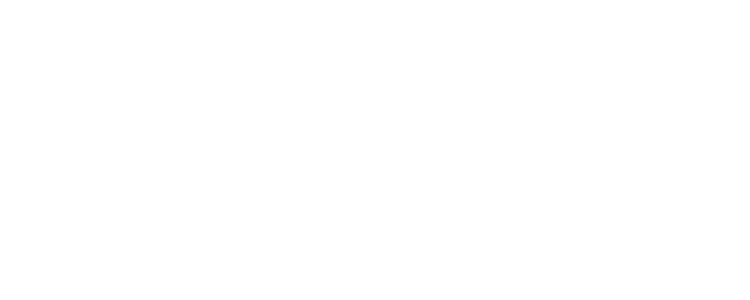}}}}\llap{\makebox[\wd1][l]{\raisebox{-0.1\wd1}{\includegraphics[width=0.36\textwidth]{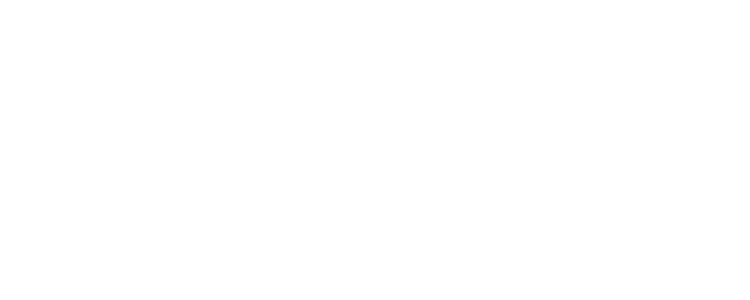}}}}
      \includegraphics[scale=0.23]{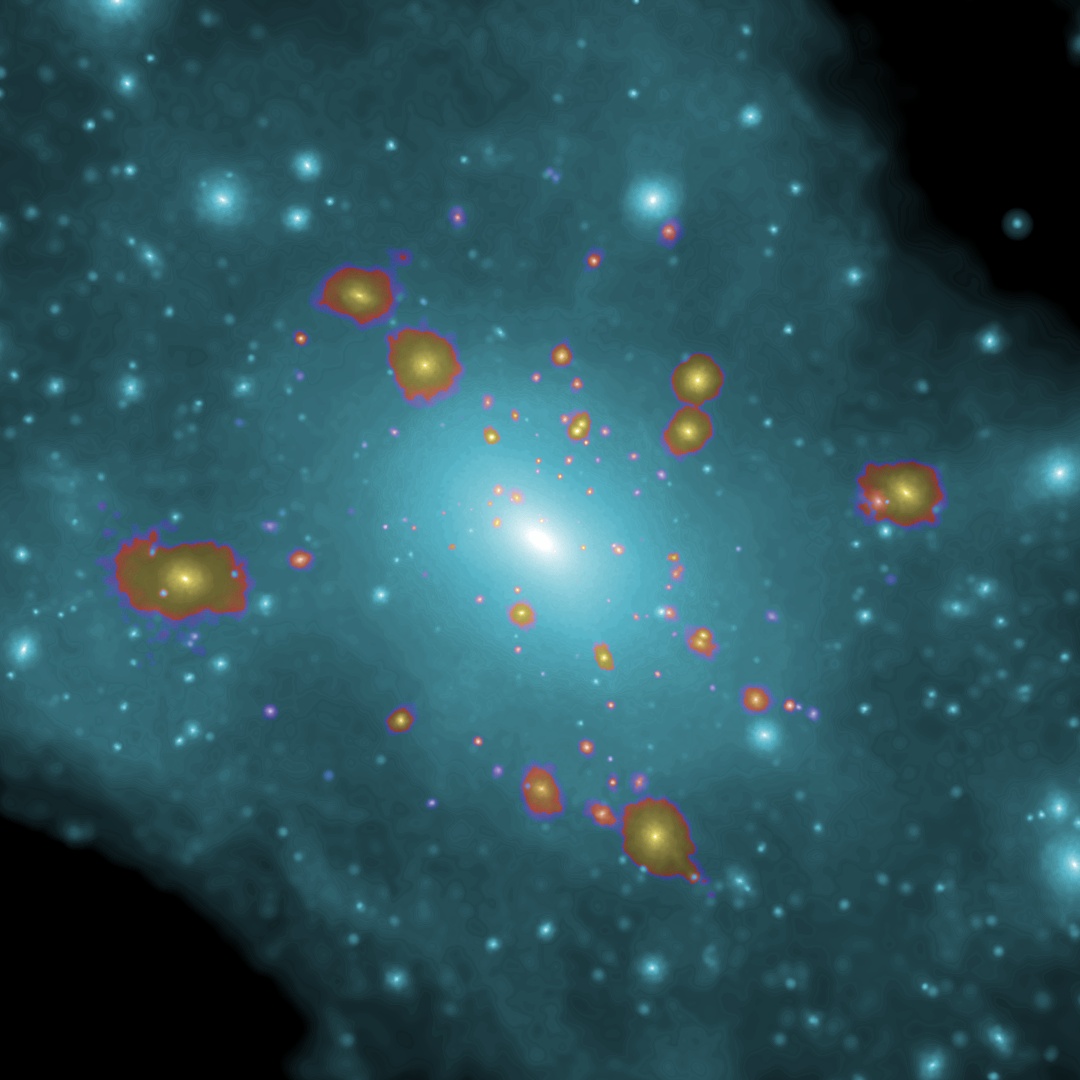}\llap{\makebox[\wd1][l]{\raisebox{0.7\wd1}{\includegraphics[width=0.36\textwidth]{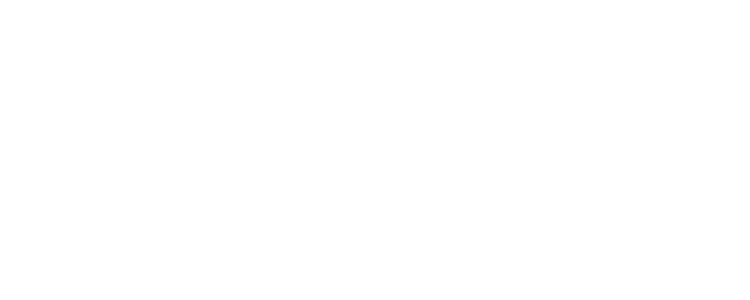}}}}
      \caption{Images of the CDM (left) and WDM (right) realizations of one of our haloes at $z=0$. Each image is $\sim450$~kpc on a side. Subhaloes within 300~kpc of the host halo centre whose progenitor was more massive at $\sim10^{8}$~$\msun$ are highlighted in red/yellow and all other particles are coloured in blue.}
    \label{fig:Imz0}
\end{figure*}

We emphasise this result with an image of one of our haloes, devised to show the distribution of subhaloes with massive progenitors. For each of the subhaloes located within 300~kpc of the host halo centre we identify its $z=0$ particles and label those particles with the subhalo progenitor's $M_\rmn{A}$. We then generate images such that particles whose progenitor subhaloes are more massive than $\sim10^{8}$~$\msun$ are highlighted in red/yellow while all other particles -- those that are members of the smooth halo, those with progenitors less massive than $10^{8}$~$\msun$, and those outside 300~kpc, are coloured in blue. We present the result in Fig.~\ref{fig:Imz0}. 

The CDM simulation shows a large number of subhaloes of various masses in the host's central region, many of which are coloured as having a massive progenitor. The outskirts of the halo are occupied by many haloes of a similar apparent size that do not have a massive progenitor (i.e. are not coloured red/yellow); therefore the $[10^{7.5},10^{8.5}]$~$\msun$ mass bracket is a combination of high-accretion mass--small radius haloes and low-accretion mass--large radius haloes. By contrast, the WDM version exhibits almost solely high infall mass subhaloes -- as indicated by the colour -- in the host centre and a paucity of similarly-sized subhaloes elsewhere; as a result the radial distribution is much steeper than for CDM. 

\begin{figure}
    \centering
    \includegraphics[scale=0.34]{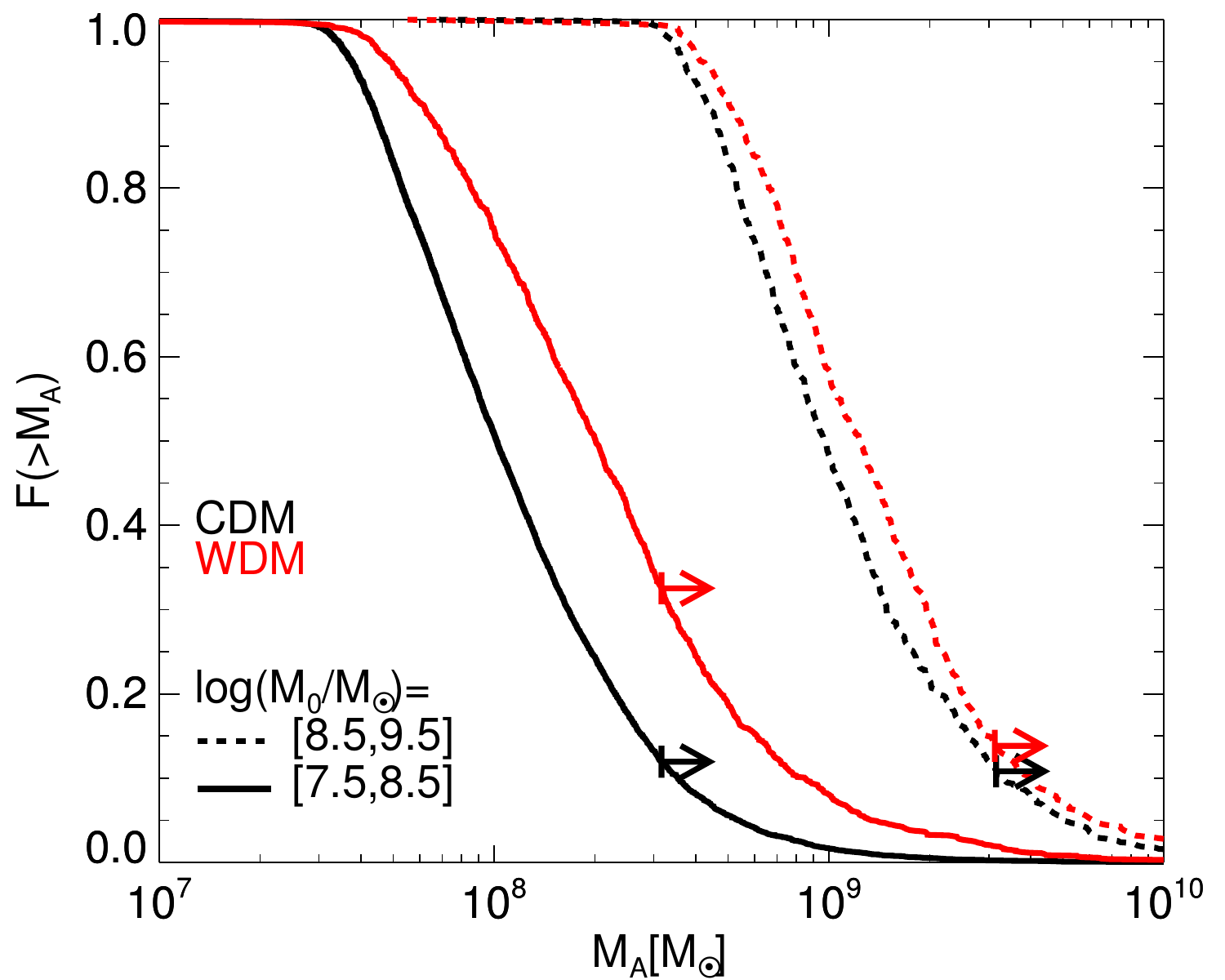}
    \caption{Normalised cumulative accretion mass functions for fixed $z=0$ mass bins. CDM data are shown in black and WDM in red. The bin with subhaloes of $M_{0}=[10^{8.5},10^{9.5}]$~$\msun$ is shown with dashed lines and that for $M_{0}=[10^{7.5},10^{8.5}]$~$\msun$ with solid lines. The arrows mark the upper edges of each $M_{0}$ bin, and therefore subhaloes to the right of the arrow have to have been stripped.}
    \label{fig:Macc8}
\end{figure}

We have argued that, to first order, dynamical friction proceeds in approximately the same manner in WDM as in CDM, i.e. that the radial distributions of CDM and WDM satellites binned by $M_\rmn{A}$ show little, if any, difference, and therefore turn to the second model feature that influences satellite distributions: the relative contributions of different infall masses to each final $z=0$ mass bin. We select subhaloes in two $M_{0}$ bins -- $[10^{7.5},10^{8.5}]$~$\msun$ and $[10^{8.5},10^{9.5}]$~$\msun$ -- and compute those subhaloes' cumulative $M_\rmn{A}$ mass functions.  We present the results in Fig.~\ref{fig:Macc8}.

In this variable we identify clear differences between the two models. The median $M_\rmn{A}$ of WDM haloes in the lower $M_{0}$ bin is a factor of two larger than CDM haloes, whereas for the higher mass bin it is instead $\sim20$~per~cent larger. 30~per~cent of WDM satellites with $M_{0}=[10^{7.5},10^{8.5}]$~$\msun$ have been stripped into the bin and 5~per~cent are the stripped remnants of $M_\rmn{A}>10^{9}$~$\msun$ progenitors, compared to only 10~per~cent and $1$~per~cent respectively of CDM haloes in this mass bracket. This compares to 15~per~cent (10~per~cent) of WDM (CDM) subhaloes stripped into the $M_{0}=[10^{8.5},10^{9.5}]$~$\msun$ bin.

\begin{figure}
    \centering
    \includegraphics[scale=0.34]{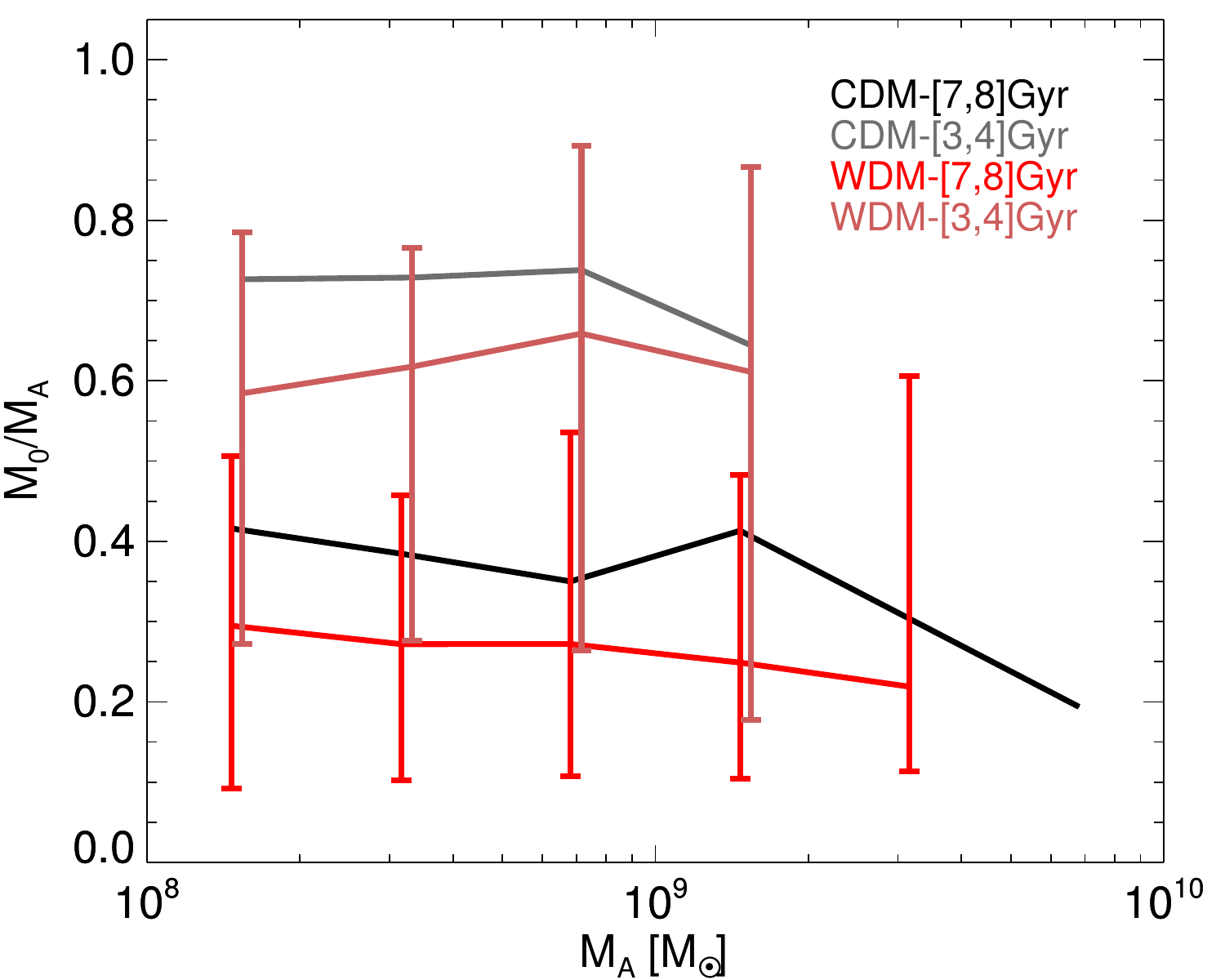}
    \caption{Ratio of present-day mass to accretion mass as a function of accretion mass. Curves are median relations, and the error bars denote the data 68~per~cent region. Results in the $[7,8]$~Gyr lookback time bin are shown in black for CDM and red for WDM, and in the $[3,4]$~Gyr lookback time bin are shown in grey for CDM and brown for WDM.  We select from all subhaloes within 300~kpc of the host centre. Error bars indicate the 68~per~cent data region, and are shown only for the WDM data~sets. }
    \label{fig:MAvsM0}
\end{figure}

In addition to the more rapid stripping time of massive subhaloes in general relative to lower mass subhaloes, we also investigate the phenomenon that WDM subhaloes are stripped more easily than their CDM counterparts with the same accretion mass due to their lower concentrations \citep[ see also][]{Bose17a}. We will consider the difference in concentration in the next subsection; here we instead do a simple test of how much the halo mass changes. We compute the ratio of present day mass, $M_{0}$, to accretion mass $M_\rmn{A}$ to obtain median stripping rates; note that we only include subhaloes that survive to $z=0$. These rates are strongly influenced by infall time: we therefore use subhaloes in two infall time bins: $t_\rmn{A}=[3,4]$~Gyr and $t_\rmn{A}=[7,8]$~Gyr. We demonstrate the differences in mass loss caused by infall time and dark matter model in Fig.~\ref{fig:MAvsM0}.

The median WDM subhalo loses approximately 15~per~cent more of its initial mass than the CDM equivalents of the same mass. This figure is roughly consistent across all masses and at both infall time bins: the median amount of mass lost in WDM (CDM) is 70~per~cent (60~per~cent) for the early infall bin and 40~per~cent (25~per~cent) in the late infall bin. 

In conclusion, we recovered the result that the CDM and WDM models predict significantly different $\sim10^{8}$~$\msun$ subhalo radial distributions. This difference arises from the sum of two contributing phenomena. First, the absence in WDM of haloes with low accretion mass and thus long dynamical friction timescales, and second, the WDM haloes experience enhanced stripping, which shifts massive, rapid-merger time subhaloes into mass bins that are lower than would be the case for CDM. 

\subsection{Radial distribution: implications for the number of stream perturbers}
\label{subsec:isp}

 A detailed analysis of the impact of subhaloes on streams requires that we take into account adiabatic contraction of the host halo \citep{Blumenthal86,Gnedin04}, the destruction of subhaloes by the MW stellar disc \citep{Ahmed17,GarrisonKimmel17,Sawala17,Richings20}, and mass loss due to evaporation of baryons from subhaloes by supernova feedback and reionisation \citep{Sawala16b}, all of which are beyond the scope of this study based on $N$-body simulations. We therefore restrict ourselves to a qualitative comparison to pure $N$-body work. 

Initial published constraints using stream gaps have made the assumption that the spatial distribution of subhaloes is the same in WDM as in CDM. While this assumption appears to work well for massive, luminous satellites \citep{Newton21}, it breaks down for dark, lower mass haloes as shown above and in \citet{Despali19b}. We investigate the implications of this result by comparing the CDM and WDM subhalo populations in the halo region where stellar stream gaps are analysed versus the population of all subhaloes out to 300~kpc.

\citet{Banik18} estimate that the GD-1 stream has a pericentre of $\sim$14~kpc and an apocentre of $\sim$30~kpc, and for Pal~5 \citep{Banik19b} estimate these parameters to be $>4$~kpc and $\sim14$~kpc respectively. Our goal is to compare the population of subhaloes in this small region of the host halo with the population of subhaloes across the host at large. For this purpose we generate subhalo samples according to distance to the host centre and present-day mass. Given the desire to obtain a reasonably sized subhalo sample, we select for our `stream-generating' subhaloes that are within 40~kpc of the centre of the host and for the host at large we select all subhaloes within 300~kpc, including those already present in the 40~kpc sample. In each radius selection we count the number of subhaloes in two $M_{0}$ mass bins -- $[10^{7},10^{8}]$~$\msun$ and $[10^{8},10^{9}]$~$\msun$ -- and compute the ratio of the number of WDM subhaloes to the number of CDM subhaloes for each CDM-WDM host pair. We also compute the total mass in subhaloes in these two radial bins and the WDM/CDM ratios for each host halo.

This calculation reflects contributions from the two differences between CDM and WDM that we have explored: enhanced stripping in WDM as shown in Fig.~\ref{fig:MAvsM0} and also the difference in the halo mass function. We analyse the contribution of these two  properties to the gaps-in-streams halo population by generating a `hybrid' data set that factors out the difference in the halo mass function as follows.

\begin{figure}
    \centering
    \includegraphics[scale=0.55]{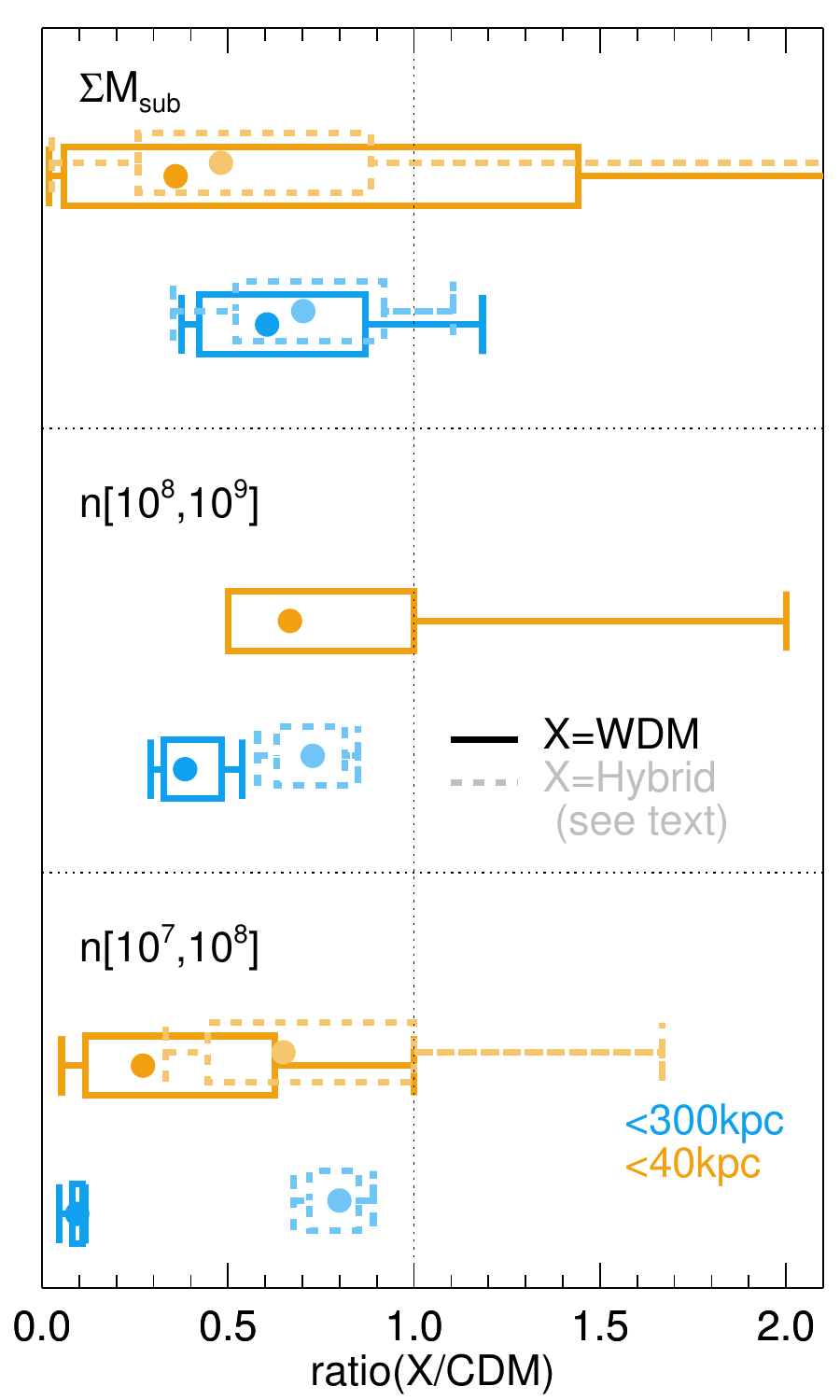}
    \caption{The ratios of substructure mass and subhalo abundance in WDM with respect to CDM for the 24 hosts. Subhalo abundance ratios in the $[10^{7},10^{8}]$ bin are indicated in the bottom panel, in the $[10^{8},10^{9}]$ bin in the middle panel, and total mass in substructure in the top panel. Results for these ratios measured for subhaloes within 300~kpc of the host centre are shown in blue, and within the inner 40~kpc in orange. For each data set the box marks 68~per~cent of hosts and the error bars mark the minimum and maximum ratio values. Median values are shown as dots. In addition to the ratios of WDM-to-CDM, shown as solid lines, we include the hybrid calculation as described in the text with faded, dashed-line boxes (top and bottom panels only).}
    \label{fig:SubFrac}
\end{figure}

For each CDM subhalo we select at random a WDM counterpart that has the same $M_\rmn{A}$ to within 50~per~cent, the same $z=0$ distance to the host centre to within 50~per~cent, and the same infall time to within 0.5~Gyr. We then regard the $M_{0}$ of the WDM counterpart as a `WDM-equivalent stripped mass'. This quantity could be thought of as the present-day mass that the CDM subhalo would have had if it had a (low-concentration) WDM density profile instead of a (high-concentration) CDM density profile. For our hybrid data~set we have the total number of CDM subhaloes but with WDM stripping rates. We therefore repeat the process discussed in the previous paragraph but replace the WDM data with our hybrid data set; however, given the small number statistics involved in the $[10^{8},10^{9}]$~$\msun$--$<40$~kpc bin we do not include hybrid data for that bin. We present the results in Fig.~\ref{fig:SubFrac}. Note that the CDM-WDM comparison contains contributions from both excess stripping and the halo mass function difference, whereas the CDM-hybrid only contain the contribution from the excess stripping\footnote{We have chosen to select counterpart subhaloes purely by accretion mass.  The predictions for individual subhaloes could be made more accurate if we selected by further variables such as eccentricity and the time of accretion; we assume that because we are assembling each host's subhaloes into a single figure that these details will not affect out result.}. 

For the full data~set, the number of WDM subhaloes in the $[10^{8},10^{9}]$ ($[10^{7},10^{8}]$) mass bins within 300~kpc of the host halo centre is 40~per~cent (10~per~cent) of the CDM number, and the scatter between haloes is a factor of 10~per~cent ($<5$~per~cent). Within the 40~kpc sphere the median suppression in the $[10^{7},10^{8}]$ bin is to 30~per~cent of CDM, so the CDM-relative subhalo abundance is three times higher than in the halo as a whole. The scatter between hosts is large, but the vast majority have higher ratios than for the $<300$~kpc bin. The $[10^{8},10^{9}]$ bin contains $\leq4$ subhaloes per host; therefore while we can report that, of the host haloes that have at least one WDM and one CDM subhalo in this bin the median number ratio is $\sim0.7$ and thus 0.4~points higher than the whole halo sample, the statistics here are poor. For the subhalo mass ratio we obtain the opposite result, with 60~per~cent WDM-to-CDM for the halo as a whole and 40~per~cent for the inner 40~kpc.    

 We can determine the relative contribution of the enhanced halo stripping in WDM to the suppression of the halo mass function by comparing these results to the hybrid model calculations. In the $<300$~kpc population the halo abundance suppression in the hybrid model result is significantly weaker than in the full WDM case -- the abundance of subhaloes is 80~per~cent of the CDM figure at $[10^{7},10^{8}]$ and 70~per~cent at $[10^{8},10^{9}]$. The median hybrid--to--CDM ratios in the $[10^{7},10^{8}]$ mass bin is 68~per~cent. Therefore, the suppression in the abundance ratios for the stream-generating subhaloes receivers a stronger contribution from stripping relative to mass function suppression than is the case for subhaloes in general. Finally, we note that the similar results are obtained for the total mass in subhaloes in the two radial populations, although this measurement is strongly influenced by the massive subhaloes and therefore the difference is much less pronounced than our two halo abundance mass bins.

We caution that in the inner region the limitations of the subhalo finder may play a role. For example, \citet{Onions12} demonstrated that {\sc subfind} may potentially underestimate the number of subhaloes near the halo centres by a factor of two. The relative abundance of CDM subhaloes in these mass bins would increase if either: i) the subhalo masses are significantly underestimated and so the cutoff in the WDM mass function effectively shifts to higher masses, or ii) there are many undetected subhaloes in the host centre that, if detected at higher resolution, would exist in larger numbers in CDM than in WDM. On the other hand, if the subhalo finder is effectively reliant on the subhalo density to make an identification, it may underdetect low-concentration WDM haloes compared to their CDM counterparts. We have investigated these possibilities in Appendix~\ref{app:ResAq} using the Aquarius $N$-body simulations of a MW-analogue halo at simulation particle masses up to a factor of 100 times smaller than is available for COCO, and we estimate from those data that the abundance of subhaloes in the $[10^{7},10^{8}]$~$\msun$ halo mass range may plausibly be underestimated by a factor of $\sim2$, in agreement with \citep{Onions12}.

One alternative approach to addressing resolution issues is to label subhaloes by a proxy for mass that is less susceptible to the uncertainties of the halo finder than is the case for the mass itself, such as the maximum of the subhalo circular velocity curve, $V_\rmn{max}$, or the mass enclosed within the radius of $V_\rmn{max}$, $M_\rmn{Vmax}$. However, these properties are also influenced by the halo density profile and its concentration, and are therefore difficult to compare between CDM and WDM. We illustrate these differences in Appendix~\ref{app:MvA}.

We now compare our results to observations. \citet{Banik21} used the structure of Pal 5 and GD-1 streams to infer the abundance of subhaloes in the $[10^{7},10^{8}]$ and $[10^{8},10^{9}]$ bins. In both mass bins they calculated that the measured value is $\sim20$~per~cent of the abundance measured in CDM simulations, and ascribe the discrepancy to gas removal and destruction by the MW disc. Our work has shown that the $N$-body prediction for 3.3~keV-WDM haloes within 40~kpc is larger than this, therefore the 3.3~keV model can explain their results provided that the impact of baryonic physics processes is less than required for CDM to match the data. We do not find an enhancement in the substructure mass fraction from $<300$~kpc to $<40$~kpc, which they measure estimate to be 14~per~cent; however, the scatter in this quantity is large and likely driven by the presence of a small number of massive subhaloes.  

In conclusion, we have demonstrated that our simulations predict that the difference between the abundances of CDM and WDM subhaloes is much smaller for the gap-generating population than for subhaloes across the rest of the halo. We subsequently infer that both CDM and WDM models with $M_\rmn{hm}\le2.8\times10^{8}$~$\msun$ are consistent with observations. However, resolution and baryonic effects are important confounding issues in the inner halo, and higher resolution hydrodynamic simulations will be required to ascertain the true ratio of CDM-to-WDM gap perturber abundances.

\subsection{Subhalo density and radial distribution}
\label{subsec:sd}

\begin{figure*}
    \centering
    \includegraphics[scale=0.35]{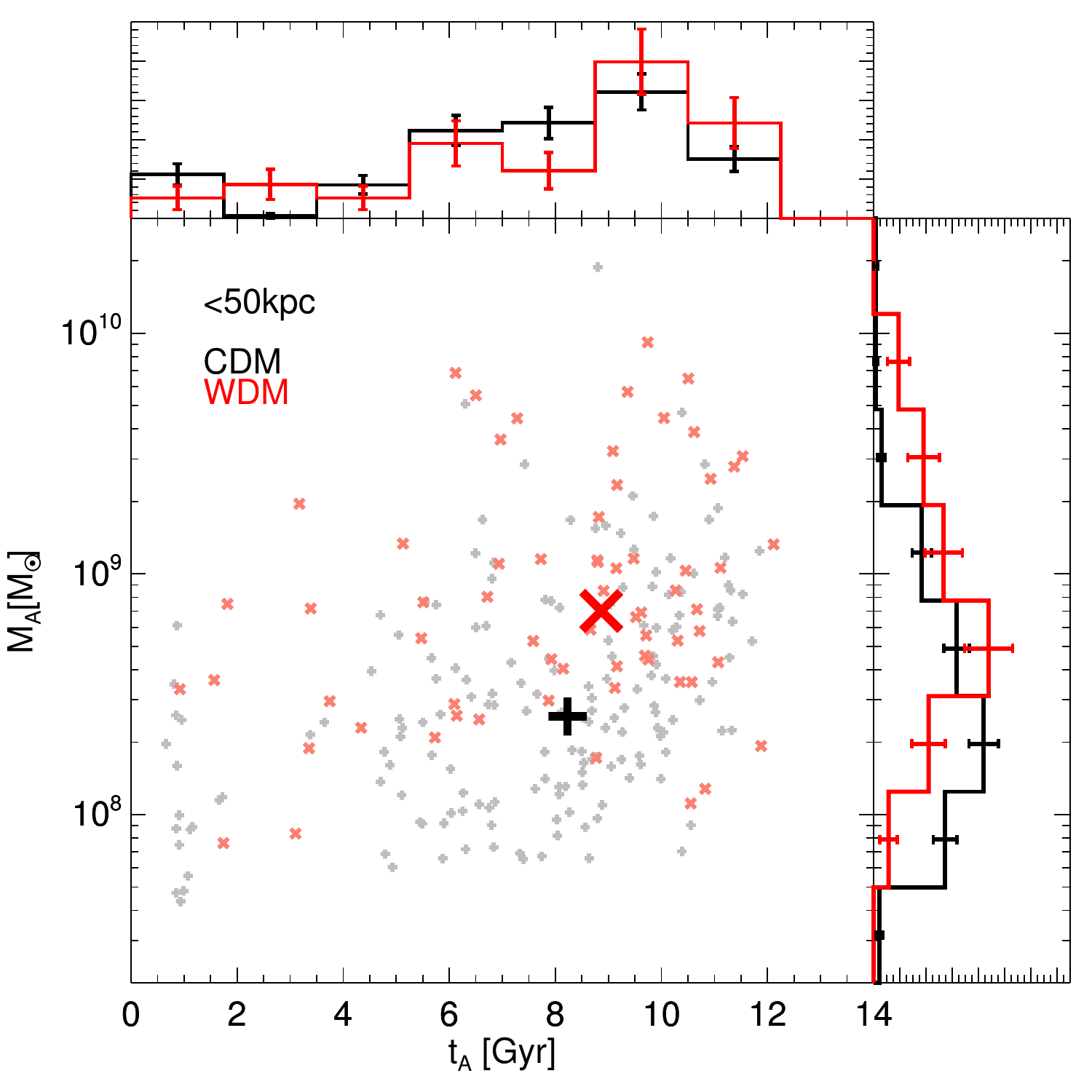}
    \includegraphics[scale=0.35]{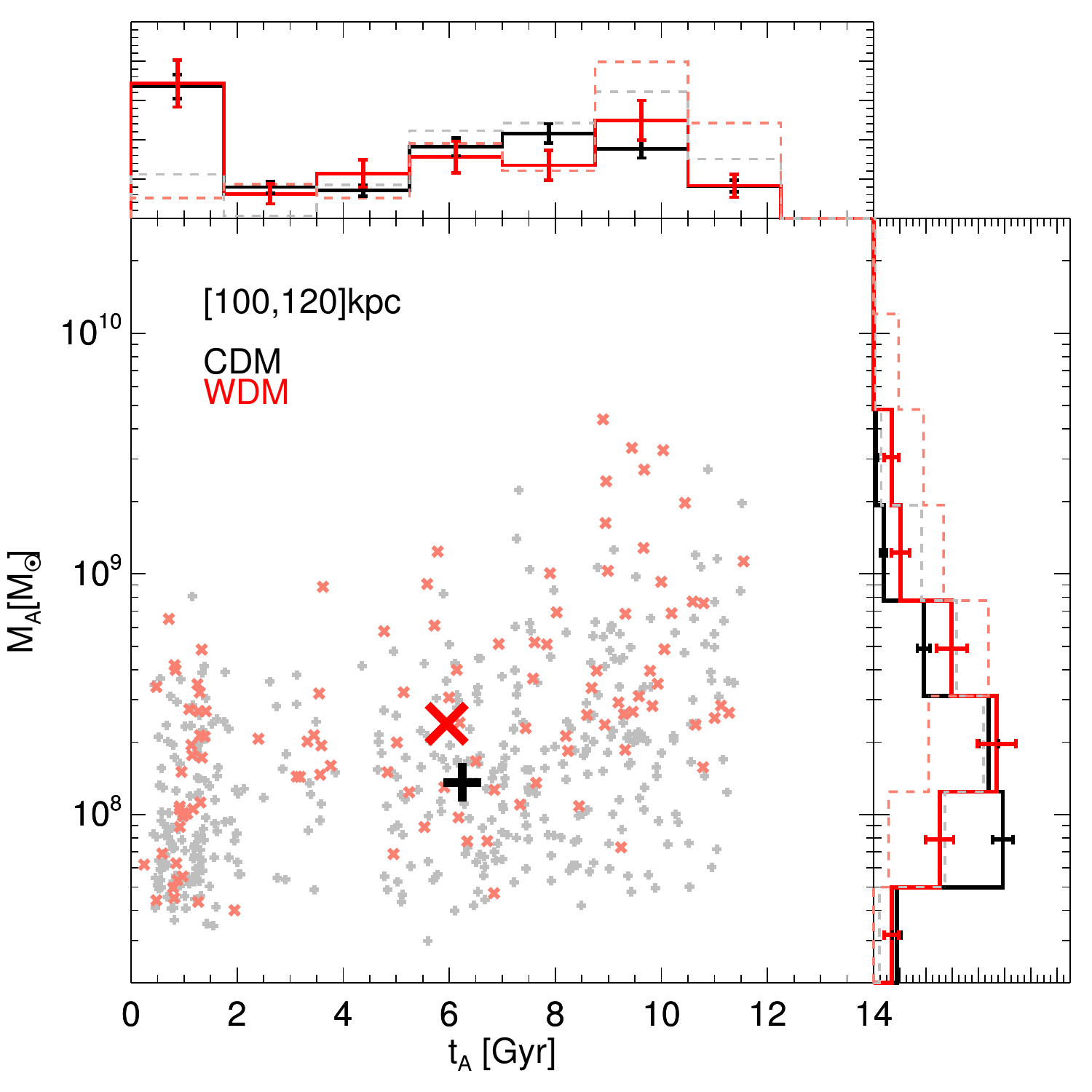}
    \caption{The distribution of accretion lookback times and accretion masses for subhaloes found today in two radial bins: within 50~kpc of the host centre (left) and the shell at [100,120]~kpc (right). CDM data are shown in black and WDM data in red. The data set medians are shown as large symbols. We include PDFs of each data set in the additional panels. In the right-hand set of panels we also reproduce the 50~kpc data PDFs as dashed lines.}
    \label{fig:MvT}
\end{figure*}

We now shift to investigating the internal properties of subhaloes found today in different radial bins, with a view to estimating the distributions of dark subhalo masses and densities. We select subhaloes with $M_{0}=[10^{7.5},10^{8.5}]$ and split them into two groups by $z=0$ radius: an inner halo sample  ($<50$~kpc from the host centre) and a halo periphery sample ($[100,120]$~kpc)  which corresponds to the mean distance to classical dwarf spheroidal satellite galaxies \citep[see ][ and references therein]{McConnachie12}. Note that here we use 50~kpc rather than 40~kpc for our central region: this is because we need better statistics to split the subhalo population by infall time as well as mass compare to mass alone. We plot $M_\rmn{A}$ as a function of accretion lookback time, $t_\rmn{A}$, for these two sets of subsamples in Fig.~\ref{fig:MvT}.

The accretion timescales for both models are very similar. At both radii there is a very small  bias towards earlier accretion times in WDM relative to CDM. The outer shell contains a population of late infalling subhaloes that is absent in the $<50$~kpc bin, and the relative size of that contribution is the same in the two models. The largest distinction between WDM and CDM is in the distribution of subhaloes accreted with masses $<10^{8}$~$\msun$ as anticipated above. The proportion of subhaloes in this mass range in WDM is suppressed by a factor of two relative to CDM in the larger radius shell but by a factor of 10 in the $<50$~kpc radius bin. There is also a strong suppression in the  relative fraction of $10^8$~$\msun$ WDM haloes within 50~kpc compared to the relative fraction within 100~kpc. Overall, the population of resolved $<50$~kpc WDM subhaloes is more biased towards high mass progenitors than either outer-radius WDM subhaloes or inner-radius CDM subhaloes.

The difference in accretion masses and accretion times between the two radial bins leads to a difference in the mass--concentration relations, because formation time and mass both correlate with concentration : concentration decreases with mass and increases with redshift \citep[][]{NFW_96,NFW_97,Neto07}. Concentration defined through the Navarro--Frenk--White profile \citep[NFW;][]{NFW_96,NFW_97} can be defined as the ratio of the profile characteristic radius to $r_{200}$ or with the characteristic overdensity, $\delta_{c}$. Another definition of characteristic radius is the half-mass radius of the gravitationally bound dark matter, $r_\rmn{half}$, as this radius is relevant in the context of gaps-in-streams impact parameters.

\begin{figure}
    \centering
    \includegraphics[scale=0.34]{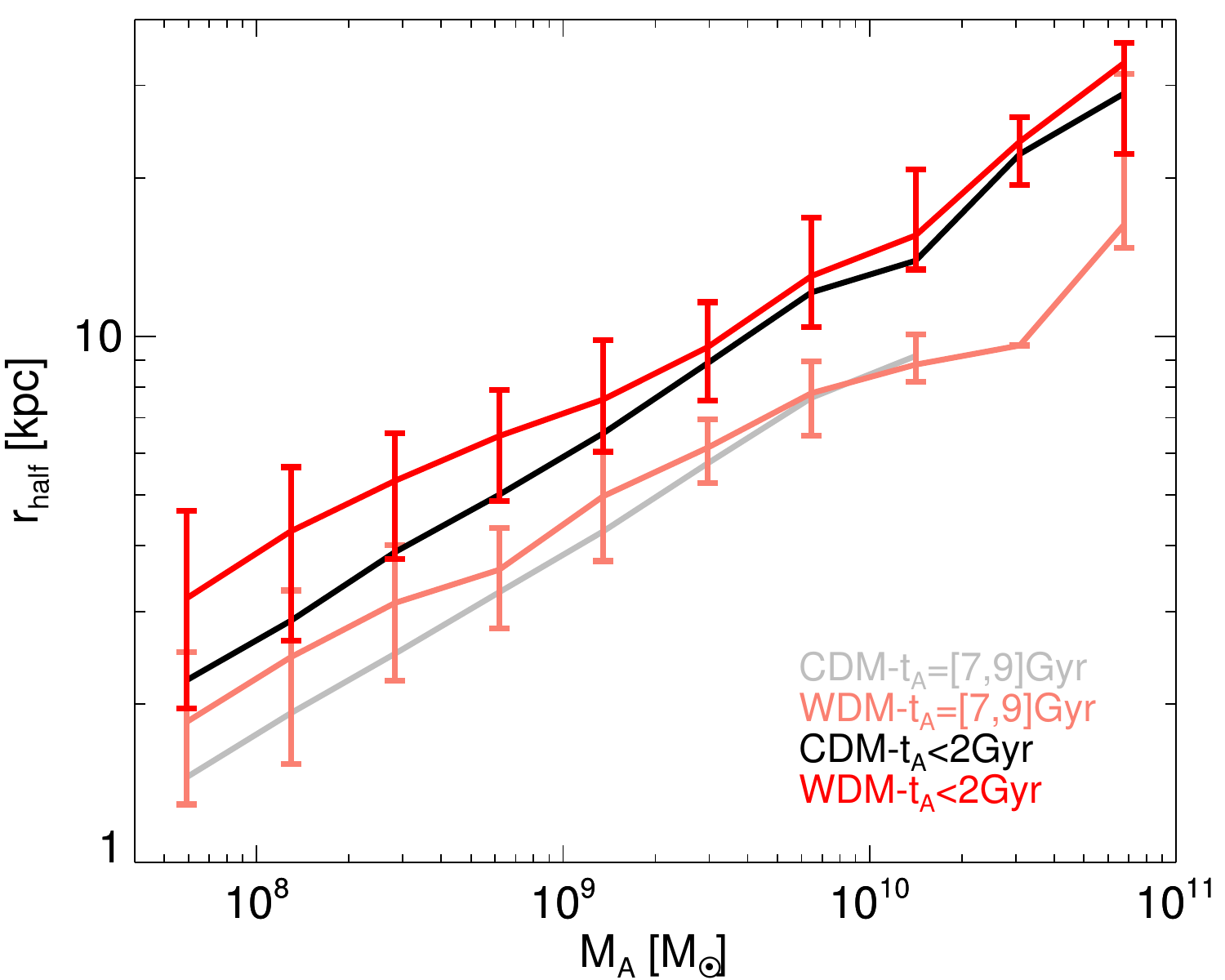}
    \caption{The distributions of half-mass radii for subhaloes at accretion, as a function of accretion mass. We show CDM and WDM data in two bins of infall time. CDM subhaloes with accretion times in the $t_\rmn{A}<2$~Gyr ($[7,9]$~Gyr) range are shown in black (grey); the WDM data in the same two lookback time bins are shown in red and pink respectively. Curves denote median relations. For the WDM data sets error bars indicate the 68~per~cent regions; the errors on the CDM data are of a similar size but are omitted for clarity purposes.}
    \label{fig:MARH}
\end{figure}

 We first compare the $r_\rmn{half}$ of CDM and WDM haloes at infall. We select an initial sample of CDM and WDM subhaloes that fall in to the host at lookback times up to $t_\rmn{A}<10$~Gyr. We split our CDM and WDM subhaloes into two distributions based on the accretion time, one at early lookback times ($[7,9]$~Gyr), and one at very low lookback times ($<2$~Gyr). Note that we use all accreted haloes in this figure, including those that are disrupted before $z=0$ or that are outside 300~kpc at $z=0$. We present the relationship between $r_\rmn{half}$ and $M_\rmn{A}$ in Fig.~\ref{fig:MARH}. 

We recover the expected result that concentrations are higher -- i.e. $r_\rmn{half}$ are smaller -- for early accretion times than late accretion times. For $M_\rmn{A}>2\times10^{9}$~$\msun$ the results for CDM and WDM are nearly identical, and both models exhibit 30~per~cent larger $r_\rmn{half}$ at late infall times. Below this threshold in $M_\rmn{A}$ the CDM $M_\rmn{A}$-$r_\rmn{half}$ distributions maintain the same power-law distribution as at higher masses. The WDM counterparts instead peel off towards larger $r_\rmn{half}$ at fixed $M_\rmn{A}$. This is a result of the inversion of the halo concentration--mass relation at low masses in \citet{Bose16a}. In the mass range $M_\rmn{A}=[10^{8},10^{9}]$~$\msun$, the early accretion WDM subhaloes exhibit a smaller $r_\rmn{half}$ / higher concentration than the late accreted CDM subhaloes in the same mass range.

We conclude this subsection by outlining possible ranges for radii of subhaloes that are relevant for gaps-in-streams studies, and in so doing demonstrate at which subhalo-stream impact parameters the difference in the mass-concentration relation between CDM and WDM subhaloes could be detected. Precise predictions would require very high resolution simulations with high time resolution output and likely 6D-phase space subhalo identification or even use a re-simulation technique similar to that of \citet{Lowing11}. In the absence of these particular requirements we restrict ourselves to an approach that arguably brackets the possible range of mass-radius distributions. Our primary interest is in the subhaloes that generate gaps in streams. These are the subhaloes located within $\sim$20~kpc of the host centre; in order to increase the sample of subhaloes we will instead select subhaloes that are as far as 50~kpc from the host centre. We will then compare these `gap-generating' subhaloes to all the subhaloes within 300~kpc.

It is in principle possible to measure the bound-mass half-light radius, $r_\rmn{half}$, of any subhalo, including those within 50~kpc. However, the halo finder can assign subhalo mass to the host halo or vice versa, and poor resolution in very small, stripped objects may affect the stripping rates. We therefore employ a method to estimate a post-tidal stripping halo mass and radius for subhaloes based on three factors: accretion mass, infall time, and a simple model of halo stripping. For each subhalo located within $<50$~kpc of the host centre we identify its accretion mass and infall time. We then compute the tidal radius, $r_\rmn{t}$, that the subhalo would have if it were situated at a distance of 20~kpc from the centre of our own MW without having its mass profile changed since infall. We use the formula for the tidal radius, $r_\rmn{t}$, presented in \citet{Springel08b}:

\begin{equation}
    r_\rmn{t} = \left[\frac{M_\rmn{sub}}{[2-d\ln M_\rmn{host}/d\ln{R}]\times M_\rmn{host}(<R)}\right]^{1/3}\times R,
\end{equation}

\noindent where $M_\rmn{sub}$ is the gravitationally bound mass, for which we will use $M_\rmn{A}$, $R$ is the distance from the subhalo centre to the host centre, for which we adopt 20~kpc, and $M_\rmn{host}(<R)$ is the spherically averaged mass profile of the host halo. For $M_\rmn{host}(<R)$ we use the total MW mass profile -- including dark matter, stars, and gas -- determined from observations by \citet{Cautun20}; therefore our result accounts for stripping due to the steepening of the dark matter halo by adiabatic contraction and also due to the presence of the stellar disc, although a full hydrodynamical treatment would be required to account for satellite destruction. We then compute the mass within $r_\rmn{t}$ to be the mass of our gaps-subhaloes, which we label $M_\rmn{t-20}$. We then compute the half mass radius for material located within $r_\rmn{t}$, and label this $r_\rmn{half-20}$. We have therefore constructed an estimate of the mass and half-mass radius of a halo shorn of its outer layers by tidal effects at 20~kpc. Note that we do not attempt to model further evolution of the halo profile \citep[see ][ for a recent discussion on stripping to small fractions of the infall mass.]{Errani21}   

\begin{figure}
    \centering
    \includegraphics[scale=0.34]{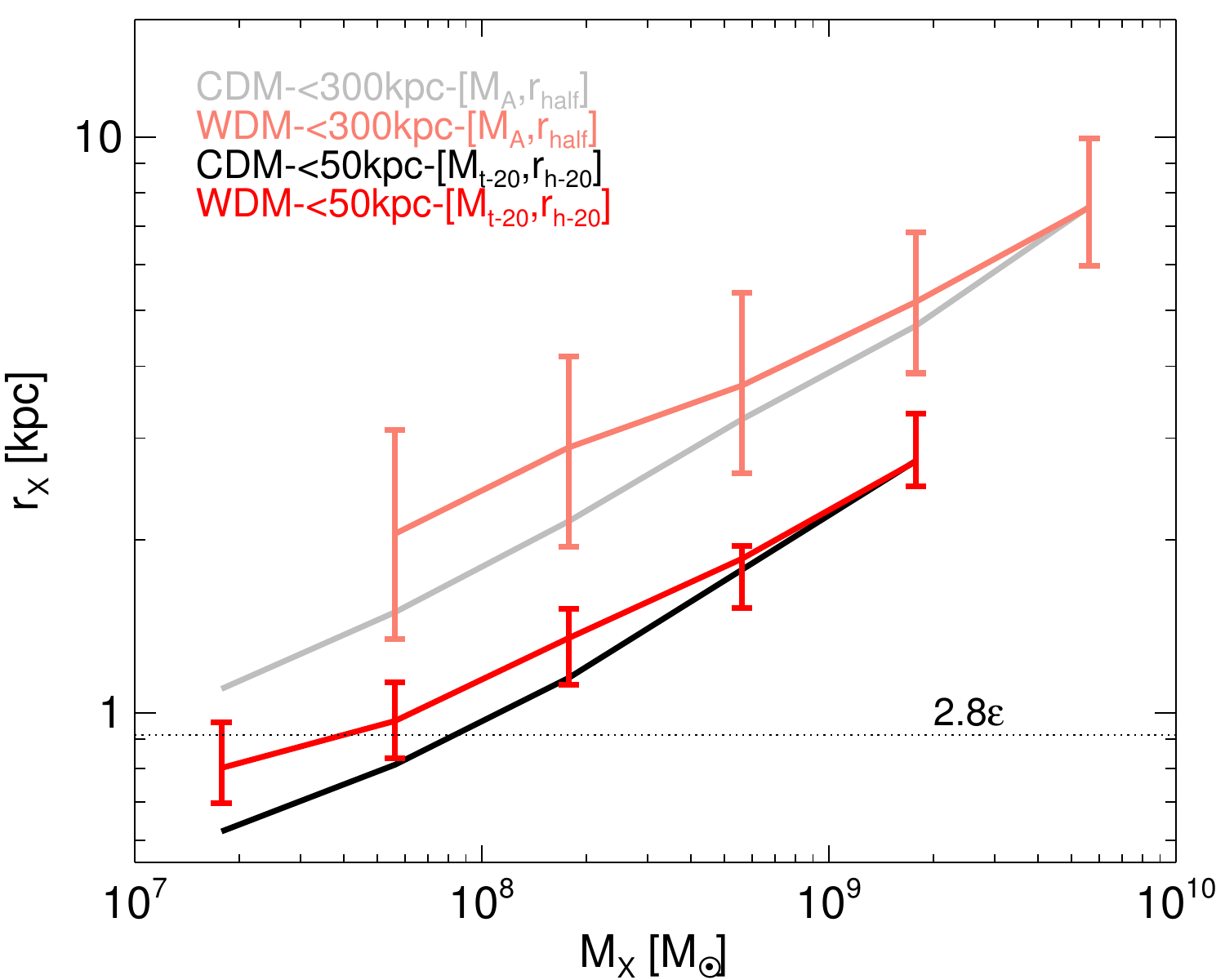}
    \caption{Approximate estimates for subhalo radii as a function of mass for CDM and WDM. We include subhaloes in two radial bins, the first within 50~kpc of the halo centre, and the second including all subhaloes out to 300~kpc. In the 50~kpc bin, we use the mass and half-mass radius definitions associated with the 20~kpc tidal stripping calculation described in the text; the CDM and WDM data for this radial bin are shown in black and red respectively. In the larger radial bin we show the half-mass radius computed at accretion as a function of the measured accretion mass, $M_\rmn{A}$; here the CDM and WDM data are shown in grey and pink respectively. Error bars mark 68~per~cent of the data, and are only shown for WDM: the CDM error bars are of roughly the same size and are omitted for clarity. The horizontal dotted line marks $2.8\times$ the gravitational softening length. The lowest-mass bin is lower than the WDM spurious subhalo mass limit $0.5M_\rmn{lim}$, and the whole-halo--$M_\rmn{A}$ bin at this mass is empty. }
    \label{fig:MTComp}
\end{figure}

Having generated a model for the gap-generating subhaloes, we turn to the whole subhalo population. For each of the subhaloes within 300~kpc we identify its $r_\rmn{half}$ at infall. The $r_\rmn{half}$  represents approximately a maximum size for each halo. We thus compute two mass-radius distributions -- $M_\rmn{t-20}$-$r_\rmn{half-20}$ for the gap-generating population and $M_\rmn{A}$-$r_\rmn{half}$ for the general population. In so doing we compare the mass--radius distributions of unprocessed subhaloes to an estimate of the mass--radius distributions of subhaloes that generate gaps. We present these in Fig.~\ref{fig:MTComp}. 

 The curves of the general population subhaloes mirror the results of Fig.~\ref{fig:MARH} closely, with both models tracking one another above a characteristic mass and diverging below it. The distribution that uses infall parameters predicts that differences between the stripping capabilities of WDM and CDM subhaloes could emerge at impact parameters as large as 4~kpc, since 16~per~cent of the WDM subhaloes  with $M_\rmn{A}\sim10^{8}$~$\msun$ have a value of $r_\rmn{half}$ greater than 4~kpc. This threshold occurs for CDM at 2~kpc, at 80~per~cent of the WDM $r_\rmn{half}$. By contrast, the $<50$~kpc population shows much more compact subhaloes at the same mass scale. The two dark matter models start to diverge from one another at tidal masses of $6\times10^{10}$~$\msun$, where the tidal median $r_\rmn{half}$ is estimated to be $<2$~kpc. We caution that this length scale is very close to the softening limit of $2.8\epsilon$. 
 
 In conclusion, we have argued that the combination of subhalo formation time, infall time, and tidal friction leads to massive, dense haloes in the regions of the MW halo where stellar stream gaps can be measured, and impact parameters of $<2$~kpc are required to discern whether perturber subhaloes are better described by CDM or WDM.

\subsection{The impact of baryonic physics: potential caveats to the {\it N}-body approach in the inner 50~kpc}
\label{subsec:barimpact}

 In the preceding sections we made predictions for the subhalo abundances in the CDM and WDM cosmologies using $N$-body simulations. Here we speculate about how baryonic physics make affect the results.

The first possible change is to the subhalo mass profiles during formation. Reionisation photons from the first stars and galaxies can evaporate gas from galaxies before stars form, which removes mass and makes the potential well shallower. This process can inhibit the accretion of more dark matter, and can lower the masses of subhaloes by an average of 30~per~cent \citep{Sawala_13}.

The interaction between this mass loss and the mass of WDM-to-CDM subhaloes in a given mass bin is complex. On the one hand, the CDM mass function is steeper than the WDM version: therefore, the relative number of subhaloes shifted {\it out of} the bin to the lower masses compared to {\it into} the bin from lower masses is larger in CDM and so the difference between the models will close slightly. On the other hand, the formation time of low mass subhaloes is delayed in WDM relative to CDM \citep{Lovell12,Lovell19a} and potentially well beyond the end of the Epoch of Reionisation, in which case a greater fraction of WDM subhaloes will have had their gas removed and so the gap between the models would widen. The lower concentrations of WDM subhaloes will also make them more susceptible to gas evacuation. Answering this question may well require bespoke simulations of reionisation to answer fully. Moreover, in satellites that form sufficiently large numbers of stars, supernova feedback could conceivably drive larger changes in the lower density WDM haloes than in CDM.

 A second consideration is the matter profile of the host halo, which we have already outlined above and approximated using the \citet{Cautun20} MW halo profile estimate. Gas cooling results in adiabatic contraction of the dark matter halo \citep{Blumenthal86,Gnedin04}, thus steepening the density gradient, shrinking the tidal radius \citep{Springel08b}, and enhancing the stripping of each satellite. The stellar disc can also play an important role in destroying satellite galaxies \citep{GarrisonKimmel17,Richings20}. The process of contraction and disc formation occurs at scales where CDM and WDM are indistinguishable, and therefore changes to the model predictions will result from differences in the subhalo mass functions and subhalo mass--concentration relations. The impact of contraction and will be strongest in the halo centre where the most massive haloes reside. It is unclear how the halo destruction rate changes between massive haloes -- which sink faster under dynamical friction and are the least concentrated -- and less massive haloes that have lower central densities over all even though their concentrations are higher. If it turns out that higher mass subhaloes are more strongly affected than low mass haloes, this will drive the difference between the models to be larger than in the $N$-body prediction. If low mass haloes are suppressed equally in CDM and WDM then the differences will instead shrink. Perhaps the most complicated outcome is the case that both low mass subhaloes are suppressed more than high mass subhaloes and low concentration (WDM) subhaloes are suppressed more than high concentration (CDM) subhaloes. It would then be down to the relative contribution of both mass and concentration to subhalo disruption as to how the final result will differ from our $N$-body predictions.

\subsection{Planes of satellites}
\label{subsec:pos}

We end our presentation of the results with a discussion of the prevalence -- or lack thereof -- of planes of satellites in CDM and WDM. It has been shown that extremely warm models, such as the 1~keV thermal relic used in \citet{Libeskind13}, can drastically reduce the number of filaments from which subhaloes can be accreted onto the host. Since filamentary accretion is an important factor in the creation of planes \citep{Libeskind05,Lovell2011}, it is possible that WDM models may generate more distinct planes than CDM and therefore explain the observed plane of satellites around the MW and other galaxies. It has been shown that the local density in which subhaloes form can affect their mass, such that WDM haloes that form in voids will be less massive than those that form in filaments \citep{Lovell19a}. The structure of filaments also changes from one model to the other, such that the spines of WDM filaments are dense enough to form stars \citep{Gao07}; we do not consider this possibility in this work.

\begin{figure*}
    \centering
        \setbox1=\hbox{\includegraphics[scale=0.23]{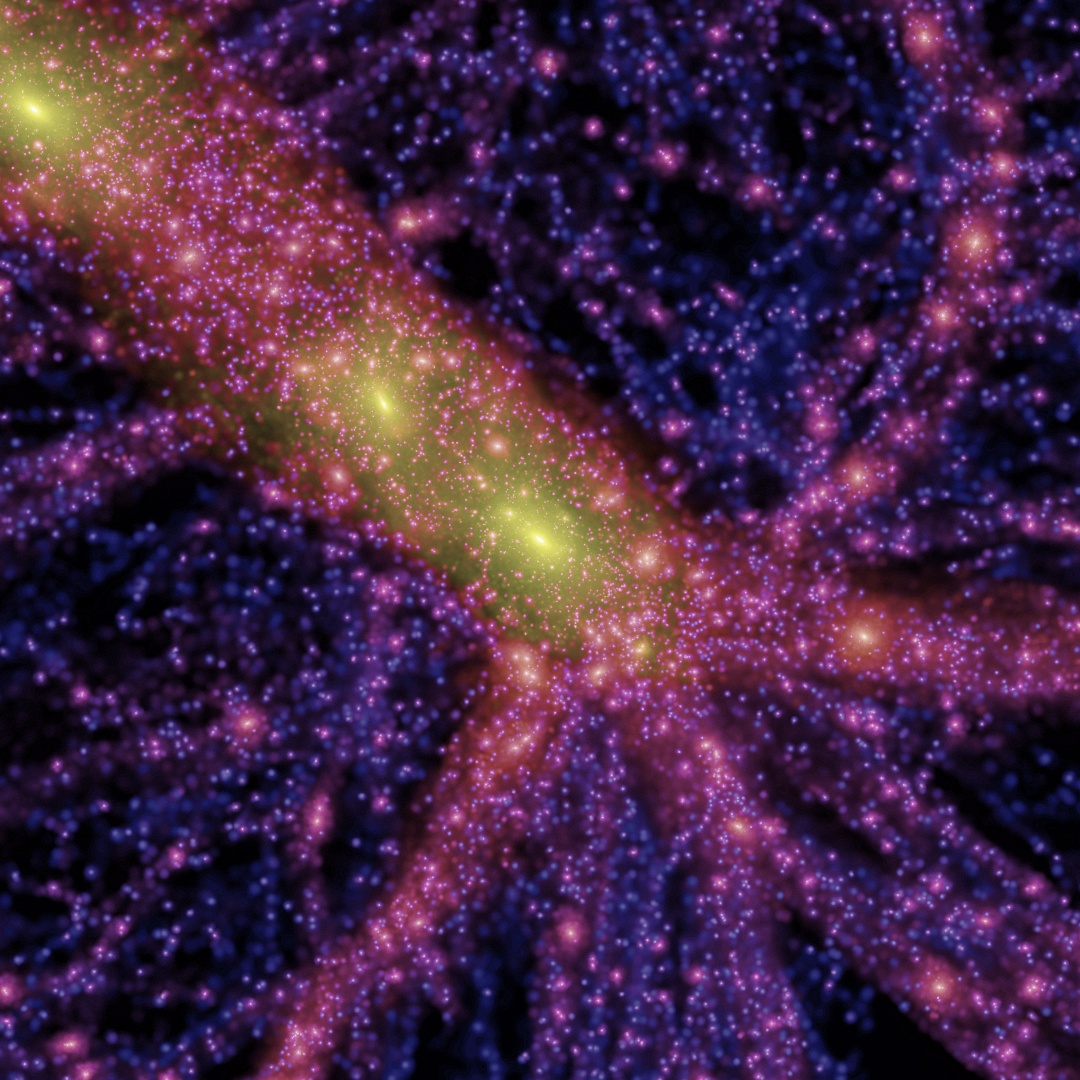}}
    
     \includegraphics[scale=0.23]{COCO_CDM_120_Bsize1000kpc_Halo006.jpg}\llap{\makebox[\wd1][l]{\raisebox{0.7\wd1}{\includegraphics[width=0.36\textwidth]{COCO_CDM_Label-eps-converted-to.pdf}}}}\llap{\makebox[\wd1][l]{\raisebox{-0.1\wd1}{\includegraphics[width=0.36\textwidth]{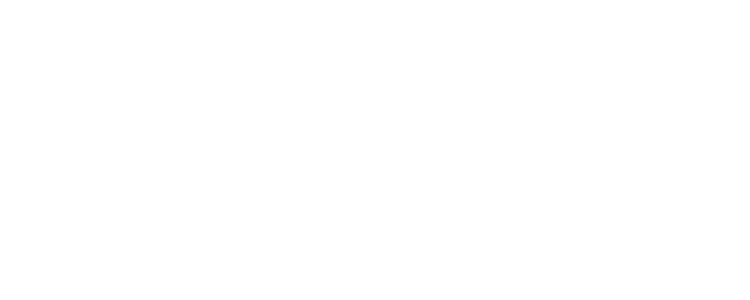}}}}
      \includegraphics[scale=0.23]{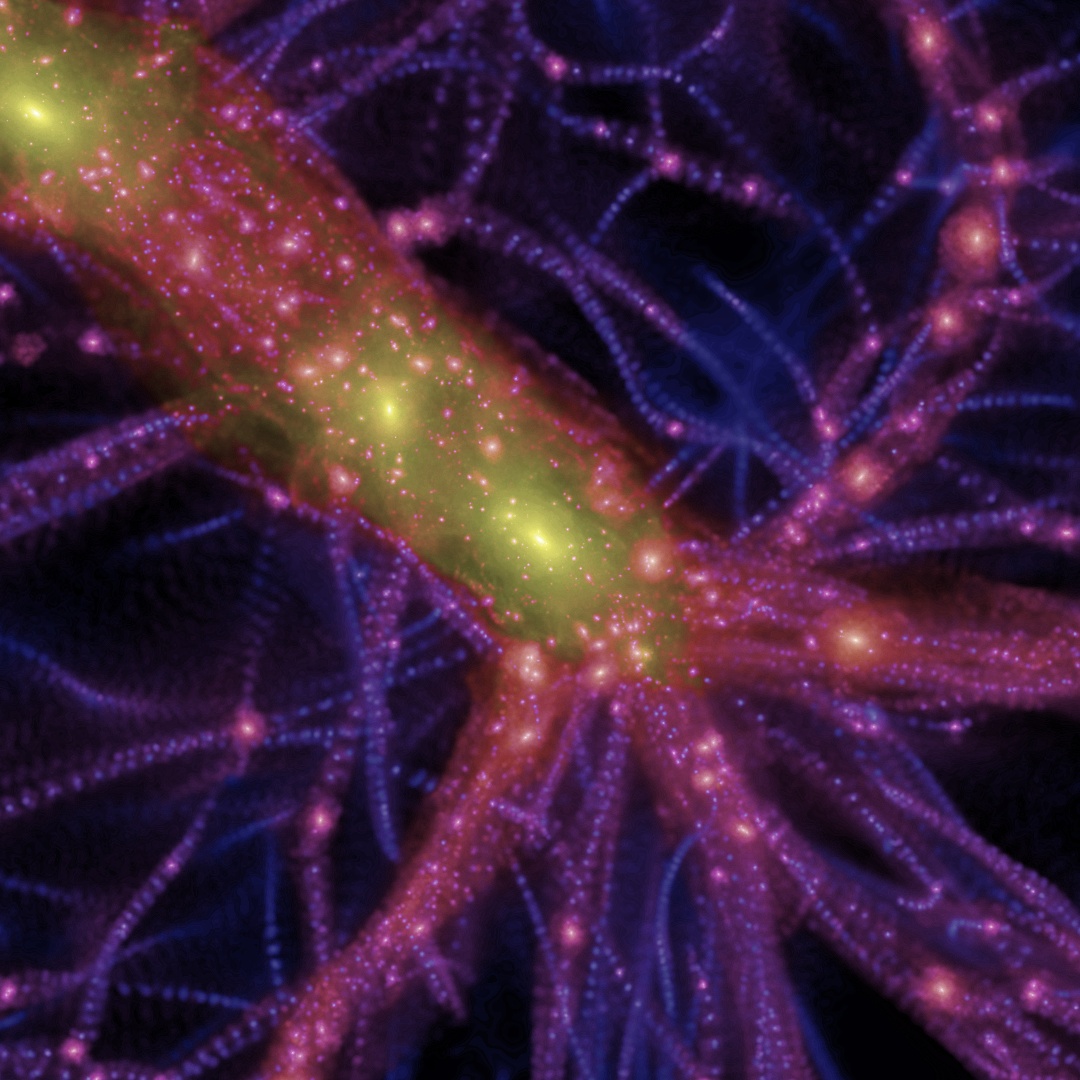}\llap{\makebox[\wd1][l]{\raisebox{0.7\wd1}{\includegraphics[width=0.36\textwidth]{COCO_WDM_Label-eps-converted-to.pdf}}}}
    \caption{Images of one host halo at $z=1.26$. CDM is shown in the left-hand panel and WDM in the right-hand panel. Each panel is 2~Mpc (comoving) on a side. The image intensity indicates the column density and the colour encodes the velocity dispersion. }
    \label{fig:Imz1p5}
\end{figure*}

In practice, the WDM cosmology includes a free parameter, $M_\rmn{hm}$, and not all observables will be affected equally at a given mass scale. For example, even though the number of dwarf haloes is strongly suppressed in WDM models with $M_\rmn{hm}\sim10^{8}$~$\msun$ \citep{Bose16a}, the large scale structure at $z=0$ is preserved \citep{Lovell20}. We will therefore investigate whether the scale of changes to the abundance of filaments in the 3.3~keV thermal relic model is sufficient to influence the existence of planes of satellites. We begin this analysis with images of the CDM and WDM counterparts of one of our haloes at $z=1.26$, shown in Fig.~\ref{fig:Imz1p5}. We will use these images to develop an initial, visual intuition about the distribution of filaments in the two models. 

The comparison of these images shows striking differences on small scales, with many more low-mass CDM haloes than WDM haloes. The main supply of subhaloes comes from a thick filament to the top left of the image plus four thinner filaments in the bottom right. Crucially, all these filaments are reproduced in the WDM version of the halo, implying that a half-mode mass $M_\rmn{hm}\le2.8\times10^{8}$~$\msun$ does not produce any striking effects on the direction of infalling haloes compared to CDM. Models with $M_\rmn{hm}$ larger than this are frequently in strong tension with observations \citep{Enzi21,Nadler21,Newton21}. Therefore it is unlikely that any viable WDM model could have a dramatic impact on the spatial distribution of satellites.

Attempts to turn this type of data into quantitative predictions for planes of satellites are numerous \citep{Lovell2011,Cautun15,Shao18}. In this study we will adopt a modified version of the \citet{Shao18} algorithm for identifying filaments and assigning subhaloes to filament membership. First, we identify the infall position of all subhaloes with peak mass ($M_\rmn{P}$) above some threshold. This position is defined as the location at which the subhalo enters the $r_{200}$ of the host halo; for subhaloes that are within 300~kpc at $z=0$ but have never entered $r_{200}$ we take the $z=0$ position. We rank these subhaloes by $M_\rmn{P}$ \footnote{We use $M_\rmn{P}$ rather than $M_\rmn{A}$ in this case because it is more likely to set the lower boundary for which subhaloes can form luminous satellite galaxies; in practice the choice of $M_\rmn{A}$ or $M_\rmn{P}$ does not affect our results.}. We identify our first filament as the location of the first ranked subhalo, and assign membership of that filament to subhaloes whose infall location is within 30~deg of that first subhalo infall position; we choose 30~deg because this is approximately the apparent size of filament at $r_{200}$ as observed from the halo centre. The second filament is then defined with the next highest $M_\rmn{P}$-ranked halo that was not assigned to the first filament, and add subhaloes to the second filament that are within 30~deg but not included in the first filament. We repeat this process until all subhaloes have been associated to a filament. We then order the filaments by the number of member subhaloes, such that the `most massive filament' is that which supplies the highest number of subhaloes, regardless of whether it contains the single most massive subhalo. Note that this method regards all subhaloes with $M_\rmn{P}$ above a chosen threshold as being equally capable of forming a galaxy and contributing to a plane; \citet{SantosSantos20} argue that the mass / luminosity of a satellite does not appear to be related to whether or not it is a member of a plane.

\begin{figure}
    \centering
    \includegraphics[scale=0.34]{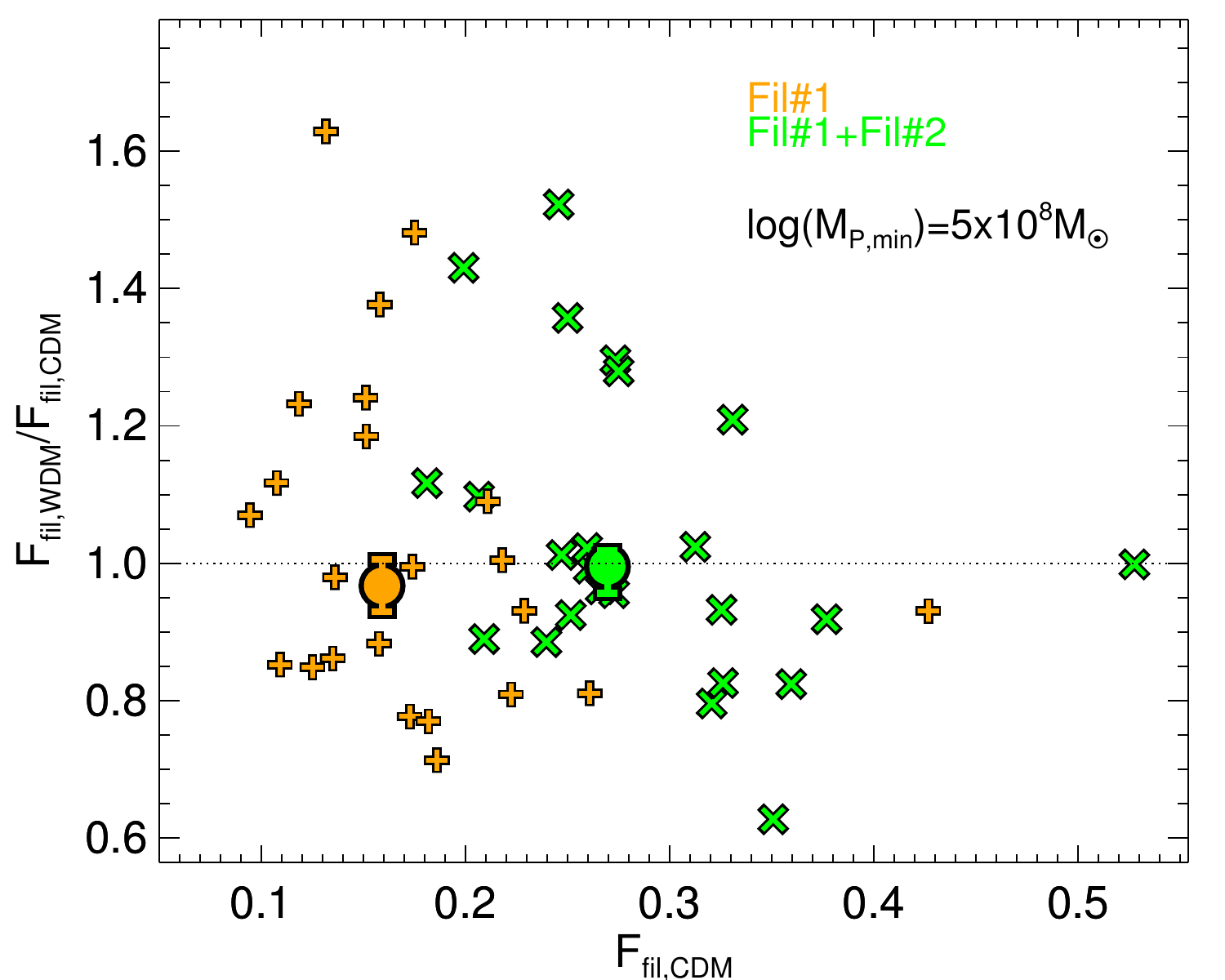}
    \caption{The fraction of subhaloes located in the first and second most massive filaments for each host as described in the text. We compute the ratio of the WDM and CDM filament fractions and plot the result as a function of the CDM counterpart fraction. We plot the filament fractions using the first filament alone as orange plus signs, and using the sum of the first and second filaments as green crosses. Filled circles denote the median values, the error bars on the circles denote bootstrap errors equivalent to 1~$\sigma$. The threshold for selecting haloes is $M_\rmn{A}>5\times10^{8}$~$\msun$.}
    \label{fig:fil}
\end{figure}

The result of this process is a list of filaments for each host, ordered by the number of member subhaloes. We perform this for a threshold in $M_\rmn{A}$ of $5\times10^{8}$~$\msun$, which we expect is the lowest halo mass capable of forming a luminous galaxy. We compute the fraction of subhaloes that are assigned to the first filament for each host, and also compute the fraction assigned to both the first and second filament. We compute the ratio of these fractions between the WDM and CDM counterparts of each host, and plot these ratios as a function of the CDM filament fraction in Fig.~\ref{fig:fil}.

The first-filament subhalo fraction ranges from 0.1 to 0.42 in CDM. The WDM first-filament fraction varies from up to 60~per~cent higher than the CDM counterpart to 30~per~cent lower, with a median fraction of $\sim5$~per~cent lower than for CDM. A similar pattern occurs when the second filament is added to the first, with the highest subhalo fraction of 0.54; the median deviation between WDM and CDM reduces to $<5$~per~cent. In both cases the bootstrap errors on the median are consistent with unity; implying that there is no statistical difference between CDM and WDM. We have repeated this process with an $M_\rmn{P}$ threshold of $3\times10^{8}$~$\msun$ and find that the results are qualitatively and quantitatively very similar to the $5\times10^{8}$~$\msun$ threshold. 

 We have included all of the subhaloes that fall into the host halo, and thus have not attempted to remove subhaloes that would pass close enough to the Galactic disc as to be disrupted. It has been argued that the removal of such subhaloes on radial orbits would increase the likelihood of satellite planes \citep{Ahmed17}. Given that these subhaloes would be the most massive -- and therefore the most alike in abundance between CDM and WDM -- there is perhaps some scope for the inclusion of baryon physics to increase the difference between the two models. A full hydrodynamical treatment will be required to check whether a disc would indeed drive the two models apart.

The main result from this analysis is that introducing the WDM cutoff changes the occupancy of filaments in a manner that is primarily stochastic rather than systematic. Any systematic effect that is present is of the order of $<10$~per~cent, and much smaller than the scatter between volumes. We have not found any evidence that attempting to restrict our subhalo selection to those subhaloes more likely to host luminous galaxies will change this finding. We therefore conclude that the impact of WDM on the filamentary accretion of satellites is small, and therefore that its influence on the planes of satellites phenomenon is at best marginal.

\section{Conclusions}
\label{sec:conc}

In this paper we have considered the spatial distribution of satellites in CDM and WDM simulations, the regime where large scale filaments and the small-scale structure and abundance of dwarf haloes combine to set subhalo properties. Within this broad topic we considered the potential for discerning between these two models using the distribution of luminous satellite galaxies, the distribution of dark subhaloes, and their implications for studies of gaps in stellar streams. We also considered the possibility that changes to filamentary accretion in WDM could increase the frequency of satellite planes.

We began by comparing the radial distributions of subhaloes in CDM and WDM, showing that the populations had different distributions when binned by mass at $z=0$ but not when binned by accretion mass. We showed that the near-defining difference between CDM and WDM -- the absence in the latter of low-mass, isolated haloes -- is the source of the difference. Low mass subhaloes in WDM are invariably the stripped remnants of high mass progenitors. At fixed present-day mass, CDM contains more subhaloes in the outer regions -- these are the largely unstripped versions of low-mass haloes either at first infall or on orbits far from the host centre. This difference is compounded by extra mass loss in WDM subhaloes due to their lower concentrations. Given that satellite galaxies will form only in the most massive subhaloes before accretion, we therefore expect that the distribution of luminous satellites will be very similar in CDM and WDM.

The less massive, dark subhaloes will have a very different distribution in the two models, with the radial distribution of WDM subhaloes being much more concentrated than that of CDM subhaloes. Therefore, the difference in abundance of WDM-to-CDM subhaloes is much reduced in the halo central regions than is the case for the halo at large. This result has important implications for the study of gaps in streams, which probe only the central $\sim30$~kpc of the MW halo. For example, the abundance of WDM-to-CDM subhaloes in the mass range $[10^{7},10^{8}]$~$\msun$ for the host at large is 10~per~cent, whereas within the inner 40~kpc the abundance of WDM subhaloes rises to 30~per~cent of CDM. We have also demonstrated that the differences within the inner 40~kpc are a combination of both extra stripping and mass function suppression, and the relative contribution of stripping is higher in the inner 40~kpc than it is for the whole halo population.

We showed that in these central region subhaloes have earlier infall times than the subhalo population in general and have more massive progenitors, and therefore these subhaloes will have a different concentration--mass relation. We derived a simple estimate for the sizes of gap-generating subhaloes, demonstrating that they are on average less than half the size of subhaloes at infall of fixed subhalo mass, and argued that subhalo--stream impact parameters of $<3$~kpc are required in order to discern whether the perturber subhalo density profile is better described by CDM or by WDM.

Finally, we considered the possibility that evidence for a cutoff in the matter power spectrum could be imprinted on the abundance of planes of satellites, with planes being more common if WDM were to erase small-scale filaments. We computed the fraction of potentially luminous subhaloes that fell onto host haloes through the first and second richest filaments, and found that there was no statistically significant difference between the CDM and WDM models. There were large stochastic differences between CDM and WDM copies of each host, but these differences were smaller than between hosts. We conclude that, if there is indeed a mass scale at which the power spectrum cutoff favours planes of satellites, it is at a thermal relic mass scale $<3.3$~keV, which is already likely ruled out by a combination of Lyman-$\alpha$ forest observations, lensing limits and MW satellite counts \citep{Enzi21,Nadler21}. We stress that our host halo selection explicitly excludes MW-analogues that have a nearby M31 companion, which may play a role in the large scale structure through which subhaloes are accreted; however, we expect that WDM will not change the likelihood that these systems generate planes of satellites.

We have thus argued that WDM models of interest have little impact on the radial distribution or the frequency of planes of luminous satellites. We have clarified the degree to which the CDM and WDM models differ in their predictions for gaps-in-streams analyses, and we have demonstrated that it is necessary to account for differences between the distribution of gap-generating subhaloes within $\sim30$~kpc of the MW centre versus the MW subhalo population at large. In order to make more precise predictions for the properties of stream gaps in both models, simulations that include stellar discs, self-consistent contraction of the host halo, baryonic processes within subhaloes, evaporation of gas by reionisation feedback, and high subhalo resolution are required.

\section*{Acknowledgements}

 MRL acknowledges support by a Grant of Excellence from the Icelandic Research Fund (grant number 206930). MC acknowledges support by the EU Horizon 2020 research and innovation programme under a Marie Sk{\l}odowska-Curie grant agreement 794474 (DancingGalaxies). CSF acknowledges support from European Research Council (ERC) Advanced Investigator grant DMIDAS (GA 786910). This work was also supported by the Consolidated Grant for Astronomy at Durham (ST/L00075X/1). WAH is supported by the
Polish National Science Center Grants No. UMO-2018/30/
E/ST9/00698 and No. UMO-2018/31/G/ST9/03388. ON acknowledges financial support from the Project IDEXLYON at the University of Lyon under the Investments for the Future Program (ANR-16-IDEX-0005) and supplementary financial support from La R\'{e}gion Auvergne-Rh\^{o}ne-Alpes. This work used the DiRAC\@Durham facility managed by the Institute for Computational Cosmology on behalf of the STFC DiRAC HPC Facility (www.dirac.ac.uk). The equipment was funded by BEIS capital funding via STFC capital grants ST/K00042X/1, ST/P002293/1 and ST/R002371/1, Durham University and STFC operations grant ST/R000832/1. DiRAC is part of the National e-Infrastructure.

\section*{Data Availability}

The data used in this paper were originally published in \citet{Hellwing16} and \citet{Bose16a}. Please the contact W. Hellwing for access.



\bibliographystyle{mnras}

\appendix
\section{Resolution study of subhalo abundance}
\label{app:ResAq}

As stated in the main text, the statistics of subhaloes close to the centre of the host are potentially affected by resolution due to the challenges of spurious disruption and the limitations of the halo finder. We check for the impact of resolution using the simulations of the Aquarius Aq-A halo \citep{Springel08b}. This halo was simulated using the WMAP-1 cosmology \citep{wmap1} and has $M_{200}=1.88\times10^{12}$~$\msun$. There are five simulations of this halo at different resolution levels, labelled from Aq-A-5 (particle mass $m_\rmn{p}=3\times10^{6}$~$\msun$) to Aq-A-1 (particle mass $1.3\times10^{3}$~$\msun$). The Aq-A-1 remains among the highest resolution simulations of a MW halo-analogue to date; the resolution of COCO is between the resolutions of Aq-A-3 and Aq-A-4. 

\begin{figure}
    \centering
    \includegraphics[scale=0.34]{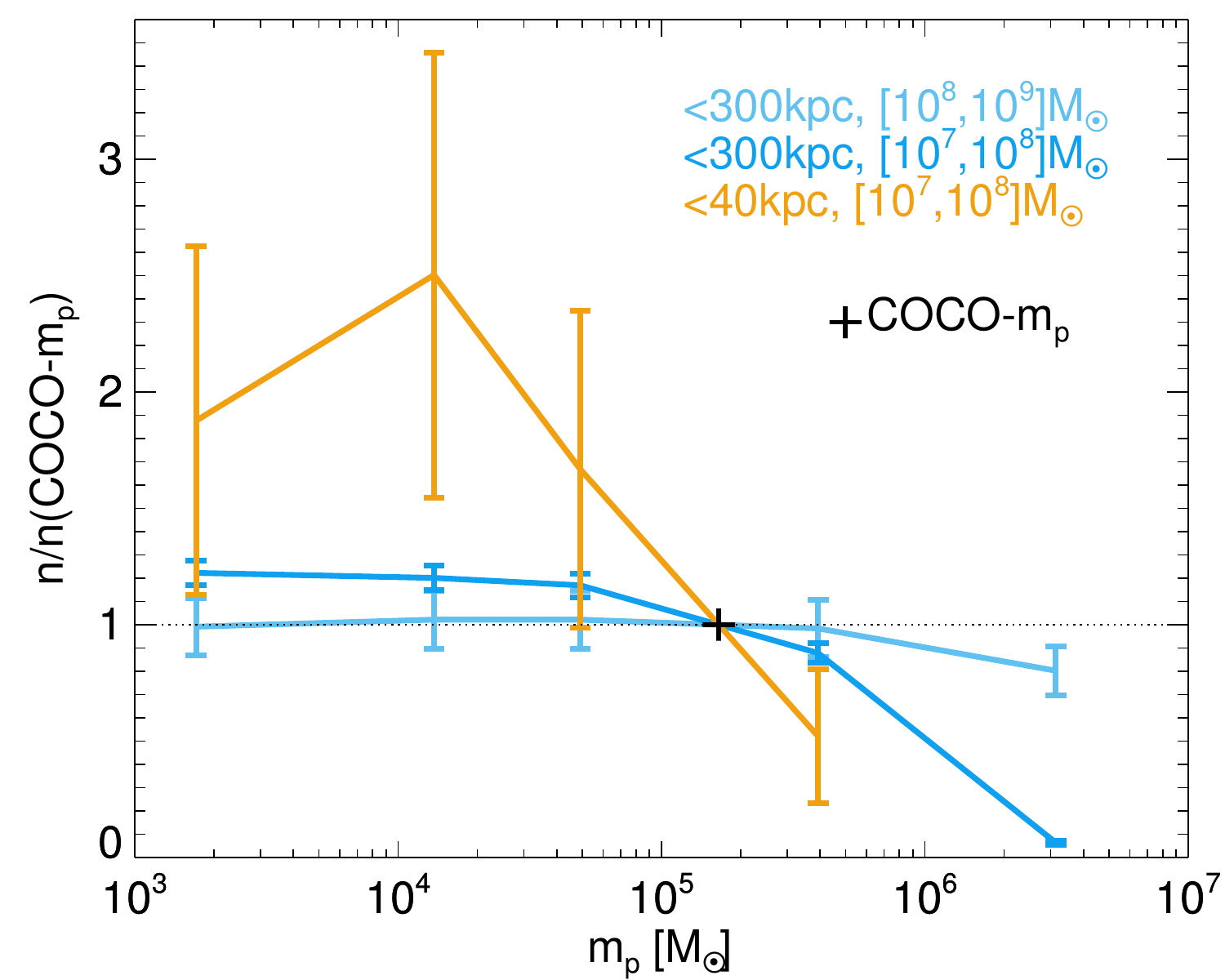}
    \caption{The subhalo abundance in mass and radial bins for the Aquarius Aq-A halo as a function of simulation particle mass. We interpolate a value for the abundance at the COCO resolution and normalise each curve by this value. Results for subhaloes $<300$~kpc from the centre and with mass $[10^{8},10^{9}]$~$\msun$ are shown in light blue, $<300$~kpc with mass $[10^{7},10^{8}]$~$\msun$ in dark blue, and $<40$~kpc with mass  $[10^{7},10^{8}]$~$\msun$ in orange. The error bars assume Poisson statistics. The COCO particle mass is shown as a black cross.}
    \label{fig:AqARes}
\end{figure}

For each of these five simulations we compute the halo abundance in radial and mass bins. For the whole halo ($<300$~kpc from the host centre) we use the mass bins $[10^{7},10^{8}]$~$\msun$ and $[10^{8},10^{9}]$~$\msun$, and for the gap-generating haloes we use $[10^{7},10^{8}]$~$\msun$; we do not use the $[10^{8},10^{9}]$~$\msun$ bin due to low statistics for this halo. We interpolate a likely value at the COCO resolution for this halo, and normalise the halo abundances by this value. We present the results as a function of $m_\rmn{p}$ in Fig.~\ref{fig:AqARes}. 

We find that the larger of the two mass bins, $[10^{8},10^{9}]$~$\msun$, is converged to better than 10~per~cent for $m_\rmn{p}<4\times10^{5}$~$\msun$ across the whole halo. The number of $[10^{7},10^{8}]$~$\msun$ subhaloes within 300~kpc is resolved with similar precision for $m_\rmn{p}<5\times10^{4}$~$\msun$, and at COCO resolution we likely underestimate the `true' number of subhaloes by 20~per~cent. At the $<40$~kpc aperture the discrepancy is clearly larger, with an average of a factor of two more subhaloes for $m_\rmn{p}<10^{4}$~$\msun$ than is inferred for COCO although the uncertainties are very large. We conclude that the difficulty of resolving subhaloes in high background density environments could well play a role on estimates of gap-generating subhalo abundance, although as we state in the main text we are not in a position to determine whether it is CDM (potentially more subhaloes to resolve) or WDM (low concentration peaks to resolve) that is more strongly affected. 

\section{Comparing accretion mass with mass proxies}
\label{app:MvA}

\begin{figure}
    \centering
    \includegraphics[scale=0.42]{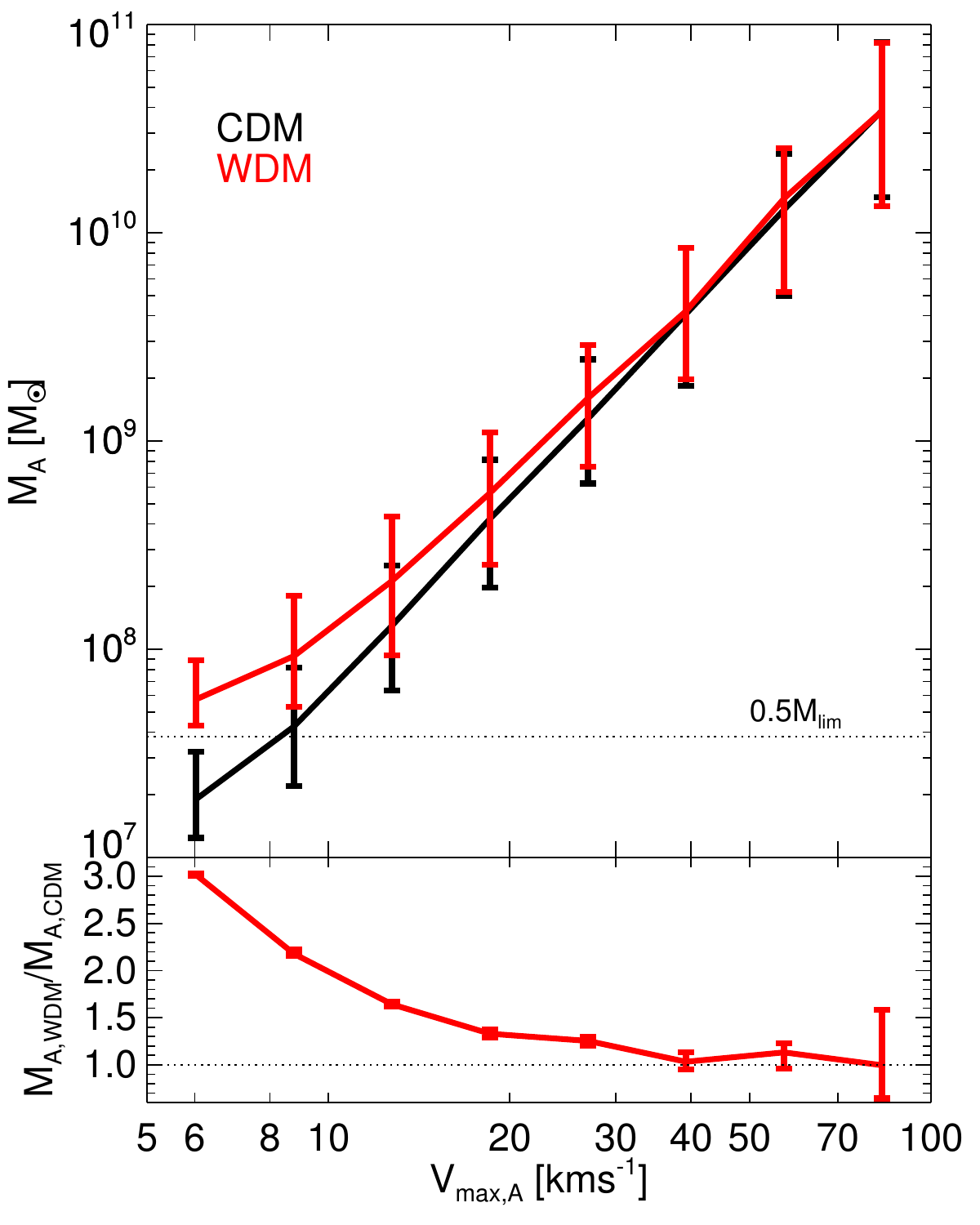}
    \caption{Top panel: the median accretion mass as a function of accretion $V_\rmn{max}$. The error bars show the 68~per~cent regions of the data; CDM is shown in black and WDM in red. The horizontal line marks the limit imposed for removing subhaloes as spurious, $0.5M_\rmn{lim}$. Bottom panel: the ratio of the WDM and CDM medians. Unlike the top panel, the bottom panel error bars are 68~per~cent bootstrap errors on the median.}
    \label{fig:VvM}
\end{figure}

 In this appendix we demonstrate the different behaviour of the mass--$V_\rmn{max}$ relation in CDM and WDM (3.3~keV thermal relic). For our sample of subhaloes at infall we compute the circular velocity curves, extract $V_\rmn{max}$ and its corresponding radius, $r_\rmn{max}$. We compute the median relation between the mass at infall, $M_\rmn{A}$ and $V_\rmn{max}$ at infall, and present the results in Fig.~\ref{fig:VvM}.

The distributions are the same for the two models at $V_\rmn{max}>40$~$\kms$. At smaller $V_\rmn{max}$ CDM and WDM diverge. At $V_\rmn{max}=20$~$\kms$ WDM haloes have $M_\rmn{A}$ 40~per~cent larger than in CDM, and at $V_\rmn{max}=10$~$\kms$ the average difference is a full factor of 2, although this latter result is partly a consequence of our spurious subhalo removal parameter, $0.5$~$M_\rmn{lim}$. This difference in mass--$V_\rmn{max}$ relations arises from the difference in the halo profiles. Starting at $M_\rmn{A}\gsim10M_\rmn{hm}$, the mass--concentration relation for WDM is the same as CDM and thus the mass--$V_\rmn{max}$ relation is also the same. As $M_\rmn{A}$ decreases, the halo concentrations of WDM haloes are systematically lower than their CDM counterparts, so $r_\rmn{max}$ increases and therefore $V_\rmn{max}$ must decrease. The inverse of this phenomenon is that for CDM and WDM haloes with the same $V_\rmn{max}$, $M_\rmn{A}$ must be higher in the WDM halo than in the CDM counterpart. We have therefore shown that care must be taken when comparing the $V_\rmn{max}$ values of subhaloes in different dark matter models.

\begin{figure}
    \centering
    \includegraphics[scale=0.42]{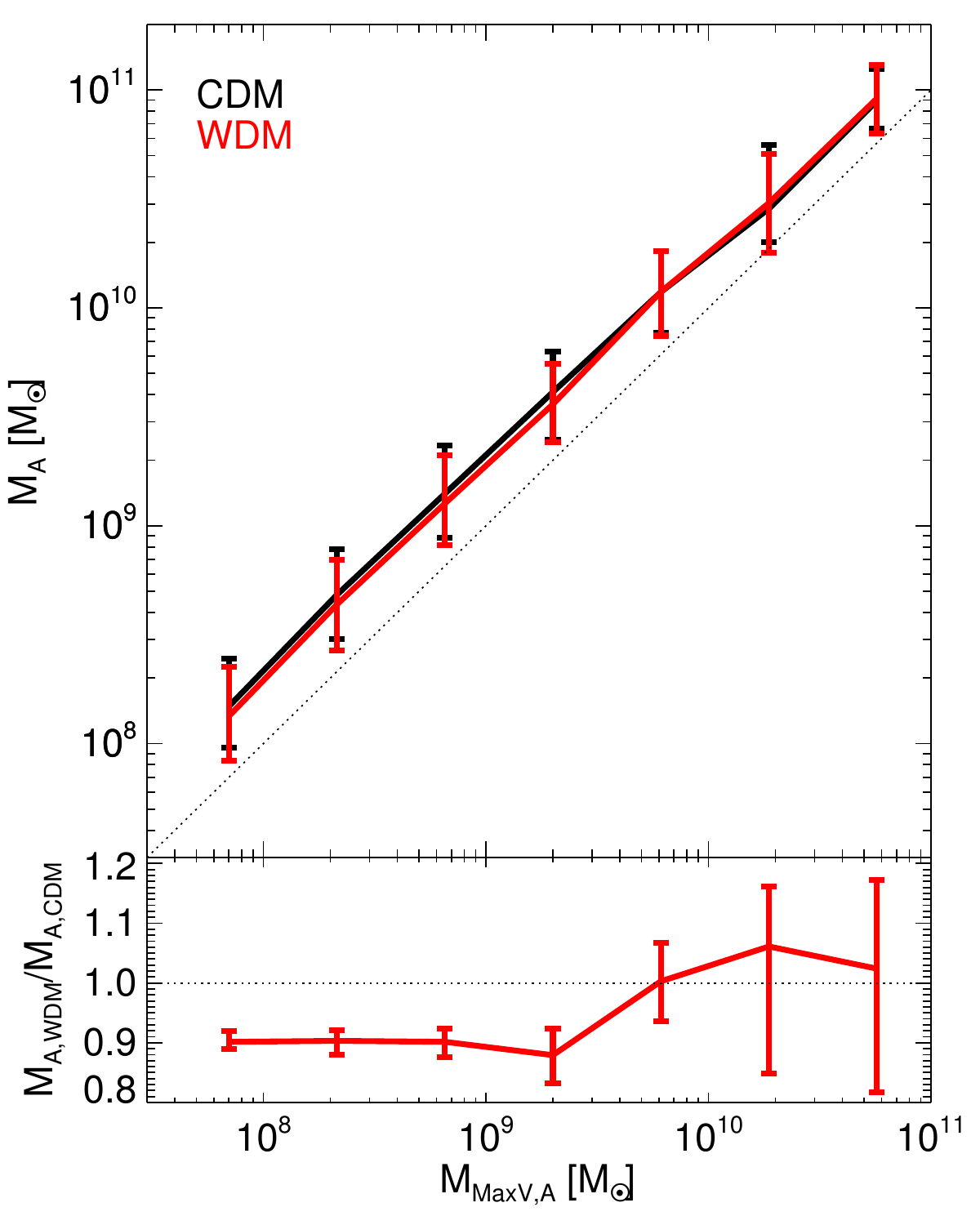}
    \caption{Top panel: the median accretion mass as a function of accretion mass within $r_\rmn{max}$, $M_\rmn{MaxV}$. The error bars show the 68~per~cent regions of the data; CDM is shown in black and WDM in red. The area below the bottom line is forbidden because $M_\rmn{MaxV}$ must be less than $M_\rmn{A}$. Bottom panel: the ratio of the WDM and CDM medians. Unlike the top panel, the bottom panel error bars are 68~per~cent bootstrap errors on the median.}
    \label{fig:MVvM}
\end{figure}

In the discussion above we have treated $M_\rmn{A}$ as approximately equal to the mass within $r_\rmn{max}$, given as  $M_\rmn{MaxV}=V_\rmn{max}^{2}\times r_\rmn{max}/G$, where $G$ is the gravitational constant. Although we have shown that the reason behind the change in $V_\rmn{max}$-$M_\rmn{A}$ relation with concentration is clear, we can also perform a direct comparison between $M_\rmn{A}$ and the computed value of $M_\rmn{MaxV}$. We present the results in Fig.~\ref{fig:MVvM}, using the same format as for Fig.~\ref{fig:VvM}. For this figure we use all subhaloes that are accreted onto the 24 hosts, irrespective of whether they survive to $z=0$. 

At $M_\rmn{MaxV}\ge6\times10^{9}$~$\msun$, the WDM and CDM median $M_\rmn{A}$ are statistically consistent with each other and are on average a factor of 2 higher than $M_\rmn{MaxV}$. For masses $<6\times10^{9}$~$\msun$ CDM and WDM masses diverge slightly such that at $M_\rmn{MaxV}\sim10^{8}$~$\msun$ the WDM version is suppressed by 10~per~cent relative to CDM. This difference is much smaller than we found for $V_\rmn{max}$. We expect that what difference there is comes from the larger WDM $r_\rmn{max}$ at fixed $M_\rmn{A}$: having a larger $r_\rmn{max}$ means that more of the total mass $M_\rmn{A}$ is included in $M_\rmn{MaxV}$.

\bsp	
\label{lastpage}
\end{document}